\documentclass[twocolumn,floatfix,a4paper,showpacs,showkeys,nofootinbib,reprint]{revtex4-1}

\usepackage{epsfig}
\usepackage{latexsym}
\usepackage{xspace}
\usepackage[colorlinks=true,linktocpage=true,linkcolor=blue,citecolor=blue,allcolors=blue]{hyperref}
\usepackage[utf8]{inputenc}
\usepackage{indentfirst}
\usepackage{enumerate}
\usepackage{color}
\usepackage{todonotes}

\usepackage[caption=false,position=top]{subfig}

\usepackage{amsmath}
\usepackage{amssymb}
\usepackage[english]{babel}
\usepackage{url}

\newcommand{\non}{\nonumber \\}

\newcommand{\eq}[1]{\begin{align} #1 \end{align}}
\newcommand{\mean}[1]{\langle #1 \rangle}

\newcommand{\sNN}{\sqrt{s_{\rm NN}}}

\begin{document}


\title{
Proton number cumulants and correlation functions in Au-Au collisions at $\sqrt{s_{\rm NN}} = 7.7-200$~GeV from hydrodynamics
}

\author{Volodymyr Vovchenko}
\affiliation{Nuclear Science Division, Lawrence Berkeley National Laboratory, 1 Cyclotron Road, Berkeley, CA 94720, USA}

\author{Chun Shen}
\affiliation{Department of Physics and Astronomy, Wayne State University, Detroit, Michigan 48201, USA}
\affiliation{RIKEN BNL Research Center, Brookhaven National Laboratory, Upton, NY 11973, USA}

\author{Volker Koch}
\affiliation{Nuclear Science Division, Lawrence Berkeley National Laboratory, 1 Cyclotron Road, Berkeley, CA 94720, USA}

\begin{abstract}
We present a dynamical description of (anti)proton number cumulants and correlation functions in central Au-Au collisions at $\sqrt{s_{\rm NN}} = 7.7-200$~GeV by utilizing viscous hydrodynamics simulations.
The cumulants of proton and baryon number are calculated in a given momentum acceptance analytically, via an appropriately extended Cooper-Frye procedure describing particlization of an interacting hadron resonance gas.
The effects of global baryon number conservation are taken into account using a generalized subensemble acceptance method.
The experimental data of the STAR Collaboration are consistent at $\sNN \gtrsim 20$~GeV with simultaneous effects of global baryon number conservation and repulsive interactions in baryon sector, the latter being in line with the behavior of baryon number susceptibilities observed in lattice QCD. 
The data at lower collision energies show possible indications for sizable attractive interactions among baryons.
The data also indicate sizable negative two-particle correlations between antiprotons that are not satisfactorily described by baryon conservation and excluded volume effects.
We also discuss differences between cumulants and correlation functions~(factorial cumulants) of (anti)proton number distribution, proton versus baryon number fluctuations, and effects of hadronic afterburner.
\end{abstract}

\maketitle


\section{Introduction}

Proton number fluctuations are one of the key observables in the beam energy scan~(BES) program at RHIC~\cite{Bzdak:2019pkr}.
The fluctuations are a sensitive probe of the QCD phase structure at finite baryon densities~\cite{Stephanov:1998dy,Stephanov:1999zu,Jeon:2000wg,Asakawa:2000wh}, the hypothetical QCD critical point in particular is thought to be very sensitive~\cite{Hatta:2003wn,Stephanov:2008qz,Luo:2017faz}. 
The STAR Collaboration has recently presented measurements of (net-)proton number cumulants up to fourth order from BES-I~\cite{Adam:2020unf,Abdallah:2021fzj}.
The measurements, which still have considerable error bars, indicate a possible non-monotonic energy dependence of the net-proton kurtosis.
It is expected that BES-II results, which will have smaller statistical uncertainties, will provide a more definitive result.
Fluctuation measurements are also being performed by other heavy-ion experiments, including the ALICE Collaboration at the LHC~\cite{Acharya:2019izy}, the NA61/SHINE Collaboration at SPS~\cite{Gazdzicki:2015ska}, and the HADES Collaboration at GSI~\cite{Adamczewski-Musch:2020slf}.

From the theory side, the heavy-ion collisions are usually described by relativistic hydrodynamics~\cite{Gale:2013da,Heinz:2013th,Romatschke:2017ejr,Shen:2020mgh}.
At a so-called particlization stage~\cite{Huovinen:2012is}, the QCD fluid is transformed into an expanding gas of hadrons and resonances, based on the picture of grand-canonical hadron resonance gas~(HRG).
Indeed, this picture works quite well to describe bulk properties of measured hadrons, such as spectra and flow, in a broad range of collision energies~\cite{Schenke:2010nt,Song:2011hk,Shen:2011eg,Karpenko:2012yf,Schenke:2020mbo}.
However, a quantitative theoretical analysis of fluctuations in this picture is challenging.
In contrast to mean hadron yields, the event-by-event fluctuations, especially the high-order cumulants, are influenced by several effects which make direct application of the grand-canonical statistical mechanics questionable.
The most important effects are the global conservation laws~\cite{Bleicher:2000ek,Begun:2006uu,Bzdak:2012an,Braun-Munzinger:2020jbk, Pratt:2020ekp} and the smearing of fluctuations due to momentum cuts~\cite{Ling:2015yau,Ohnishi:2016bdf}.
Other mechanisms include volume fluctuations~\cite{Gorenstein:2011vq,Skokov:2012ds,Braun-Munzinger:2016yjz} or diffusion in the hadronic phase~\cite{Steinheimer:2016cir,Asakawa:2019kek}.

The two main issues mentioned above were recently addressed in Ref.~\cite{Vovchenko:2020kwg} at LHC energies via a generalized Cooper-Frye procedure called the subensemble sampler, utilizing the approximately boost invariant nature of heavy-ion collisions at the LHC parameterized by the blast-wave model.
In this work we extend this method to the RHIC-BES energies.
This is achieved in the following way.
First, we relax the assumption of boost invariance.
Instead, we employ realistic particlization hypersurfaces obtained from (3+1)-dimensional viscous hydrodynamics simulations using code MUSIC~\cite{Paquet:2015lta, Shen:2017bsr, Shen:2020jwv}.
Second, we calculate the cumulants of (anti)proton number distribution emitted from the hypersurface subject to global baryon conservation analytically~(rather than using Monte Carlo as in Ref.~\cite{Vovchenko:2020kwg}).
For this purpose we use a recently developed subensemble acceptance method 2.0~(SAM-2.0)~\cite{SAM2p0}, which allows one to perform a correction of cumulants of accepted protons for the effect of exact global baryon conservation analytically.
As a result, we are able to calculate cumulants of (anti)proton number distribution emerging from hydrodynamics to high orders without the need to generate large numbers of Monte Carlo events.
The results are then confronted with the experimental data of the STAR Collaboration.

The paper is organized as follows.
Section~\ref{sec:cumu} presents the method to calculate cumulants of proton and baryon number distribution at particlization.
The calculation results for Au-Au collisions at RHIC-BES energies are presented in Sec.~\ref{sec:results}.
Discussion and summary in \ref{sec:summary} close the article.

\section{Cumulants from hydrodynamics}
\label{sec:cumu}

We employ relativistic viscous hydrodynamics to simulate the evolution of a system created in 0-5\% Au-Au collisions at RHIC, using the open-source code \texttt{MUSIC v3.0}~\cite{Schenke:2010nt,Paquet:2015lta,Denicol:2018wdp}.
We perform hydrodynamic simulations with event averaged initial density profiles at each collision energy, which describes the expansion of quark-gluon plasma created in the earlier stage of the collision and its transition to a hadron gas.
Cumulants of proton and baryon number distributions are computed at the end of hydrodynamic evolution at particlization.

\subsection{Hydrodynamics}

The 3D initial conditions are taken from Ref.~\cite{Shen:2020jwv}.
They are based on the Glauber collision geometry, employ local energy and momentum conservation, and calibrated to reproduce the measured proton transverse momentum distributions and midrapidity yields at different collision energies.
This makes it suitable for the analysis of second and higher-order proton cumulants, which we perform here.

The hydrodynamic evolution is calculated in \texttt{MUSIC v3.0} by numerically solving the equations corresponding to energy-momentum and baryon number conservation, as well as Israel-Stewart-type relaxation equation describing the viscous stress tensor.
We include shear viscous corrections but neglect bulk viscous corrections and baryon diffusion.
We employ a NEOS-BSQ\footnote{NEOS-BSQ stands for nuclear equation of state with baryon number, strangeness, and electric charge} equation of state from Ref.~\cite{Monnai:2019hkn} which interpolates between lattice QCD equation at large temperatures~\cite{Bazavov:2014pvz}, described via the Taylor expansion, and hadron resonance gas at low temperatures.
This equation of state imposes vanishing net strangeness, $n_S = 0$, and a fixed charge-to-baryon ratio, $n_Q/n_B = 0.4$, in every fluid cell.
The shear viscosity to entropy baryon ratio is temperature- and chemical-potential-dependent, the details can be found in Fig.~4 of Ref.~\cite{Shen:2020jwv}.
The hydrodynamic equations are solved in Milne coordinates, $(\tau,\,x,\,y,\,\eta_s)$, and evolved in $\tau$ until all computational cells reach a threshold energy density $\varepsilon_{\rm sw}$ for particlization.

\subsection{Cumulants of baryon-proton number distribution at particlization}

The hydrodynamic evolution ends at a particlization hypersurface $\sigma(x)$ of constant “switching” energy density $\varepsilon_{\rm sw}$. 
The value of $\varepsilon_{\rm sw} = 0.26$~GeV/fm$^3$ has been adjusted in Ref.~\cite{Shen:2020jwv} to fit bulk observables and it is used here throughout.
Note that a further improvement of the proton spectra at $\sqrt{s_{\rm NN}} \gtrsim 39$~GeV can be achieved by increasing the value of $\varepsilon_{\rm sw}$ at those energies~\cite{Oliinychenko:2020znl}.
In our calculations we observed a mild effect of a changing $\varepsilon_{\rm sw}$ on proton cumulants~(this is detailed in the Appendix~\ref{app:esw}), mainly in a form of stronger excluded volume effects at larger densities, thus we keep $\varepsilon_{\rm sw} = 0.26$~GeV/fm$^3$ at all energies for consistency.

The QCD fluid is transformed at particlization into a system consisting of hadrons and resonances.
The momentum spectrum for hadron species $j$ emerging from hydrodynamics is given by the Cooper-Frye formula~\cite{Cooper:1974mv}\footnote{We neglect the shear viscous corrections to particle momentum distributions at particlization, which only has a small influence on the high-$p_T$ tail of the distribution~\cite{Shen:2010uy}.}
\eq{\label{eq:CF}
\omega_p \frac{d N_j}{d^3 p} = \int_{\sigma(x)} d \sigma_\mu (x) \, p^\mu \, f_j[u^\mu(x) p_\mu;T(x),\mu_j(x)],
}
with
\eq{
f_j[u^\mu p_\mu;T(x),\mu(x)] = \frac{d_j \, \lambda_j(x)}{(2\pi)^3} \, \exp\left[ \frac{\mu_j(x) - u^\mu(x) p_\mu}{T(x)} \right].
}

Here $\lambda_j(x)$ describes deviations from the ideal gas distribution function due to interactions.
We assume that these deviations are momentum-independent.
In the ideal HRG limit one has $\lambda_j(x) = 1$.
On the other hand, this factor is different from unity in case of a non-ideal HRG.
Here we employ the excluded volume HRG~(EV-HRG) model with repulsive baryon-baryon interactions~\cite{Vovchenko:2016rkn,Vovchenko:2017xad}, which has been observed to provide an improved description of the lattice QCD data on baryon number susceptibilities near the pseudocritical temperature $T_{\rm pc} \sim 155$~MeV at $\mu_B = 0$ compared to the standard ideal HRG.
This model has been used in our previous study of proton fluctuations for the LHC energies~\cite{Vovchenko:2020kwg} and we refer to that work for further details on the EV-HRG model.
We also perform calculations in the ideal HRG limit to establish the relevance of baryon repulsion in the investigated observables.
We use the open source package \texttt{Thermal-FIST v1.3} in all our HRG model calculations~\cite{Vovchenko:2019pjl}.

The particlization hypersurfaces, consisting of a list of fluid elements each containing the normal four-vector $d \sigma_\mu$, the fluid four-velocity $u^\mu$ as well as energy and baryon densities, are available via~\cite{MUSICinput}.
For each hypersurface element we recalculate the values of the temperature $T(x)$ and the chemical potential $\mu_{B}(x)$ such that the HRG model equation of state at these $T$-$\mu_B$ values matches the energy and baryon densities corresponding to this hypersurface element. We also enforce $n_Q/n_B = 0.4$ and $n_S = 0$ for each fluid element to determine the electric charge and strangeness chemical potentials $\mu_Q$ and $\mu_S$.

With the numerical output from \texttt{MUSIC} the Cooper-Frye integral becomes a sum over all fluid elements:
\eq{\label{eq:CFnum}
\omega_p \frac{d N_j}{d^3 p} = \sum_{i \in \sigma} \delta \sigma_\mu (x_i) \, p^\mu \, f[u^\nu(x_i) p_\nu;T(x_i),\mu_j(x_i)].
}

In what follows we neglect the modification of the shape of (anti)baryon spectra due to resonance decays and evolution in the hadronic phase.
All baryons are modeled as thermal particles with nucleon mass $m_N = 0.938$~GeV emitted from a Cooper-Frye hypersurface.
These simplifications make it feasible to evaluate proton number cumulants analytically.
We relax these assumptions in Appendix~\ref{app:MC} using a Monte Carlo approach and show that they have only small influence on the resulting proton number cumulants, at least up to the third order.
Equation~\eqref{eq:CFnum} is sufficient to calculate the number of (anti)baryons in a given momentum acceptance by integrating over the momenta.
To calculate fluctuations, however, we need a generalization beyond the standard Cooper-Frye prescription.

Let us first calculate the cumulants in the grand-canonical limit, i.e. neglecting the exact global conservation of the baryon number.
We shall correct the cumulants for the baryon number conservation via the recently developed method of Ref.~\cite{SAM2p0} afterwards.
We further assume that the dynamical correlation length $\xi$ that defines the range of correlations is smaller than any other relevant scale, such that one can assume $\xi \to 0$.
This is in analogy to the model of critical fluctuations at freeze-out studied in Refs.~\cite{Ling:2015yau,Jiang:2015hri}.
In our case, where particle number correlations in the grand-canonical ensemble are attributed purely to the excluded volume effect, the emission of particles from all the hypersurface elements proceeds independently.
To calculate the cumulants of (anti)baryon number distribution inside a particular momentum acceptance it is thus sufficient to sum up contributions from all the volume elements independently.
The number of (anti)baryons emitted from a hypersurface element $i$ fluctuates in accordance with the grand-canonical susceptibilities $\chi^{B^\pm}$ of (anti)baryon number fluctuations. The corresponding cumulants, therefore, read
\eq{
\delta \kappa_n^{B^\pm, \rm gce} (x_i) = \delta V_{i}^{\rm eff} \, [T(x_i)]^3 \, \chi^{B^\pm}_n (x_i).
}
Here $\delta V_{i}^{\rm eff} = \delta \sigma_\mu (x_i) \, u^\mu (x_i)$ is the effective volume of a hypersurface element $i$.
The probability $p_{\rm acc}(x_i;\Delta p_{\rm acc})$ that an (anti)baryon emitted from a fluid element $i$ ends up in a momentum acceptance $\Delta p_{\rm acc}$ can be calculated from the Cooper-Frye formula~\eqref{eq:CFnum}:
\eq{
& p_{\rm acc}(x_i;\Delta p_{\rm acc}) =  \non 
& \frac{\displaystyle \int_{p \in \Delta p_{\rm acc}} \frac{d^3 p}{\omega_p} \delta \sigma_\mu (x_i) \, p^\mu \, f[u^\nu(x_i) p_\nu;T(x_i),\mu_j(x_i)]}{\displaystyle \int \frac{d^3 p}{\omega_p} \delta \sigma_\mu (x_i) \, p^\mu \, f[u^\nu(x_i) p_\nu;T(x_i),\mu_j(x_i)] }~.
}

The contribution $\delta \kappa_n^{B^\pm, \rm gce} (x_i; \Delta p_{\rm acc})$ of element $i$ to the $n$th order cumulant of the accepted (anti)baryons is obtained by convoluting the cumulants $\{ \delta \kappa_l^{B^\pm, \rm gce} (x_i) \}$, $l = 1\ldots n$ with the binomial distribution with probability $p_{\rm acc}(x_i;\Delta p_{\rm acc})$.
One obtains~(see e.g.~\cite{Savchuk:2019xfg}):
\eq{\label{eq:kbino}
& \delta \kappa_n^{B^\pm, \rm gce}(x_i; \Delta p_{\rm acc}) = \non 
& \qquad \sum_{l=1}^n \, \delta \kappa_l^{B^\pm, \rm gce} (x_i) \,
B_{n,l}\left(\phi'_t, \ldots, \phi^{(n-l+1)}_t \right)~.
}
Here $\phi \equiv \phi(t,p_{\rm acc}) = \ln(1- p_{\rm acc}+e^t p_{\rm acc})$ and $B_{n,l}$ are partial Bell polynomials.

In the same way we can also obtain the (anti)baryon cumulants $\delta \kappa_n^{B^\pm, \rm gce}(x_i; \overline{\Delta p_{\rm acc}})$ outside the acceptance $\Delta p_{\rm acc}$, i.e. for $p \not\in \Delta p_{\rm acc}$. This is achieved by substituting $p_{\rm acc}(x_i;\Delta p_{\rm acc}) \to 1 - p_{\rm acc}(x_i;\Delta p_{\rm acc})$ in Eq.~\eqref{eq:kbino}.

To obtain (anti)proton number cumulants one can apply the arguments of Kitazawa and Asakawa~\cite{Kitazawa:2011wh,Kitazawa:2012at}: based on the isospin randomization in the hadronic phase, the (anti)proton cumulants are obtained by the binomial filtering of the (anti)baryon cumulants.
Note that this argument does not necessarily require a long-lasting hadronic phase with many pion-nucleon scatterings to randomize the isospin.
The binomial filtering is valid also in the case of models where primordial correlations between baryons do not depend on the isospin. 
This is the case for the EV-HRG model studied here, where the same excluded volume parameter is used for all baryon pairs, regardless of their isospins.
The probability that a randomly chosen (anti)baryon is (anti)proton is simply the ratio between the mean numbers of (anti)protons and (anti)baryons, $q(x_i) = \frac{\mean{N_{p^\pm}(x_i)}}{\mean{N_{B^{\pm}}(x_i)}}$.
The value of $q(x_i)$ is calculated using the HRG model in each hypersurface element. 
To be consistent with the experimental conditions realized in the STAR experiment, we include all weak decay contributions~\cite{Oliinychenko:2020znl}.
In practice, this yields $q(x_i) \approx 1/2$ in most cases.

We shall use the binomial distribution argument to obtain the joint baryon-proton cumulants $\delta \kappa^{B^\pm,p^\pm}_{n,m}(x_i;\Delta p_{\rm acc})$ in terms of baryon cumulants $\delta \kappa_n^{B^\pm, \rm gce}(x_i; \Delta p_{\rm acc})$ and the proton-to-baryon ratio $q(x_i)$, for each hypersurface element $x_i$.
We calculate the joint cumulants because these quantities will later be needed to apply the correction for baryon number conservation using the method of Ref.~\cite{SAM2p0}.
Given the probability $P(N_B)$ to have $N_B$ baryons the joint probability $P(N_B,N_p)$ to have $N_B$ baryons and $N_p$ protons is
\eq{
P(N_B,N_p) = B(N_B,N_p;q) \, P(N_B),
}
where
\eq{
B(N_B,N_p;q) = \binom{N_B}{N_p} \, q^{N_p} (1-q)^{N_B-N_p}
}
is the binomial distribution.

The joint cumulant generating function for $N_B$ and $N_p$ reads
\eq{
G(t_B,t_p) & = \ln \mean{e^{t_B N_B} e^{t_p N_p}} \non
& = \sum_{N_B} P(N_B) e^{t_B N_B} \sum_{N_p} \, B(N_B,N_p;q) \, e^{t_p N_p} \non
& = \sum_{N_B} P(N_B) e^{\gamma(t_B,t_p,q) \, N_B} \non
& = G_B[\gamma(t_B,t_p,q)]~.
}
Here $G_B$ is the cumulant generating function of the $N_B$ distribution and
\eq{\label{eq:gammbino}
\gamma(t_B,t_p,q) = t_B + \ln[1-(1-e^{t_p})q]~.
}
To obtain Eq.~\eqref{eq:gammbino} we used an identity $\sum_{N_p = 0}^{N_B} B(N_B,N_p;q) \, e^{t_p N_p} = (1 - q + e^t \, q)^{N_B}$.

The joint cumulants $\kappa^{B^+,p^+}_{n,m}$ of baryon-proton distribution correspond to the Taylor expansion coefficients of the generating function $G(t_B,t_p)$ around $t_B = t_p = 0$.
The corresponding derivatives of $G(t_B,t_p)$ are evaluated with the help of Fa\`a di Bruno's formula and expressed in terms of the cumulants $\kappa_n^B$ of the $N_B$ distribution:
\eq{\label{eq:pB}
\kappa^{B^+,p^+}_{n,m} & = \kappa_n^{B^+}, \quad m = 0~,\\
\kappa^{B^+,p^+}_{n,m} & = \sum_{k=1}^m \, \kappa_{n+k}^{B^+} \, B_{m,k}(\gamma'_{t_p}, \ldots, \gamma^{(m-k+1)}_{t_p}), \qquad m \geq 1~.
}

The same procedure applies for the joint cumulants of antiproton-antibaryon distribution.
Rewriting Eq.~\eqref{eq:pB} for the cumulants corresponding to the accepted particles emitted from volume element $x_i$ we get
\eq{
& \delta \kappa^{B^\pm, p^\pm, \rm gce}_{n,m}(x_i; \Delta p_{\rm acc})  = \delta_{m,0} \, \delta \kappa_n^{B^\pm, \rm gce}(x_i; \Delta p_{\rm acc}) \non
& \quad + \sum_{k=1}^m \, \delta \kappa_{n+k}^{B^\pm, \rm gce}(x_i; \Delta p_{\rm acc}) \, B_{m,k}(\gamma'_{t_p}, \ldots, \gamma^{(m-k+1)}_{t_p})~.
}

The joint cumulants of all accepted (anti)baryons/(anti)protons are obtained by summing over the contributions of all the hypersurface elements:
\eq{\label{eq:kappanBgce}
\kappa^{B^\pm, p^\pm, \rm gce}_{n,m}(\Delta p_{\rm acc}) = \sum_{i \in \sigma} \, \delta \kappa^{B^\pm, p^\pm, \rm gce}_{n,m}(x_i; \Delta p_{\rm acc})~.
}

The joint net-baryon/net-proton cumulants can be obtained straightforwardly in the case when there are no grand-canonical correlations between baryons and antibaryons.
This is the case for the EV-HRG model used here.
The cumulants read:
\eq{
\kappa^{B, p, \rm gce}_{n,m}(\Delta p_{\rm acc}) & = \kappa^{B^+, p^+, \rm gce}_{n,m}(\Delta p_{\rm acc}) \non
& \qquad + (-1)^{n+m} \, \kappa^{B^-, p^-, \rm gce}_{n,m}(\Delta p_{\rm acc})~.
}
We also list here, for completeness, the joint net-baryon/proton and net-baryon/antiproton cumulants
\eq{
\kappa^{B, p^+, \rm gce}_{n,m}(\Delta p_{\rm acc}) & = \kappa^{B^+, p^+, \rm gce}_{n,m}(\Delta p_{\rm acc}) \non
& \qquad +
\delta_{m,0} \, (-1)^n \, \kappa^{B^-, p^-, \rm gce}_{n,0}(\Delta p_{\rm acc})~, \\
\kappa^{B, p^-, \rm gce}_{n,m}(\Delta p_{\rm acc}) & = \delta_{m,0} \, \kappa^{B^+, p^+, \rm gce}_{n,0}(\Delta p_{\rm acc}) \non
& \qquad +
(-1)^n \, \kappa^{B^-, p^-, \rm gce}_{n,m}(\Delta p_{\rm acc})~,
}

The joint net-baryon/(net-)(anti-)proton cumulants outside the acceptance are obtained in the same fashion, by substituting $p_{\rm acc}(x_i;\Delta p_{\rm acc}) \to 1 - p_{\rm acc}(x_i;\Delta p_{\rm acc})$.

\subsection{Correction for global baryon conservation}

To account for the exact global baryon conservation we apply a generalized version of the subensemble acceptance method~(SAM) developed in Ref.~\cite{SAM2p0}.
The SAM-2.0 allows one to calculate the effect of global conservation of a conserved quantity, such as net baryon number, on the cumulants of arbitrary non-conserved quantity, such as (net) proton number.
The original SAM~\cite{Vovchenko:2020tsr,Vovchenko:2020gne} was formulated for uniform thermal systems in the thermodynamic limit and acceptances in the coordinate space.
The SAM-2.0 extends the method to non-uniform systems and momentum space acceptances. 
The method takes joint net-baryon/(net-)(anti-)proton number cumulants calculated inside and outside the acceptance without the account of exact baryon conservation and produces the cumulants that are subject to global baryon conservation, i.e. it provides a mapping
\eq{\label{eq:SAM2po}
\kappa^{B, p, \rm ce}_{n,m}(\Delta p_{\rm acc}) = \tilde{\mathcal{S}}\left[ \kappa^{B, p, \rm gce}_{n,m}(\Delta p_{\rm acc}),  \kappa^{B, p, \rm gce}_{n,m}(\overline{\Delta p_{\rm acc}})  \right]. 
}
Here $\overline{\Delta p_{\rm acc}}$ corresponds to particles outside the momentum acceptance $\Delta p_{\rm acc}$.
The details of the procedure to calculate the mapping $\tilde{\mathcal{S}}$ are presented in Ref.~\cite{SAM2p0}.
The grand-canonical cumulants $\kappa^{B, p, \rm gce}_{n,m}(\Delta p_{\rm acc})$ and  $\kappa^{B, p, \rm gce}_{n,m}(\overline{\Delta p_{\rm acc}})$ entering the right-hand-side of Eq.~\eqref{eq:SAM2po} were calculated in the previous subsection.

It is assumed in SAM-2.0 that the system is sufficiently large, such that the means of all the relevant quantities correspond to the maximum of probability distribution~(see Ref.~\cite{SAM2p0} for details).
This assumption is realized in central Au-Au collisions at RHIC-BES.
The second assumption of the method is that the distributions inside and outside the acceptance are independent, i.e. that cumulants $\kappa^{B, p, \rm gce}_{n,m}(\Delta p_{\rm acc})$ and $\kappa^{B, p, \rm gce}_{n,m}(\overline{\Delta p_{\rm acc}})$ are additive, such that $\kappa^{B, p, \rm gce}_{n,m} = \kappa^{B, p, \rm gce}_{n,m}(\Delta p_{\rm acc}) + \kappa^{B, p, \rm gce}_{n,m}(\overline{\Delta p_{\rm acc}})$.
This assumption is satisfied exactly in the ideal HRG model and to a high accuracy within the EV-HRG model at RHIC-BES energies.
As discussed in Ref.~\cite{SAM2p0}, even in an extreme case where the additivity of cumulants does not hold, the SAM-2.0 results exhibit only small deviations from the exact result, thus the possible deviations from the true results for the EV-HRG model applications considered in the present work are expected to be negligible.
The results presented in the next section that incorporate the effect of global baryon conservation have all been obtained using SAM-2.0.
A \texttt{Mathematica} notebook which calculates the mapping $\tilde{\mathcal{S}}$ is available via~\cite{SAMgithub} and used in this work.

\section{Results}
\label{sec:results}

Calculations are performed for 0-5\% Au-Au collisions for collision energies $\sNN = 7.7$, $14.5$, $19.6$, $27$, $39$, $62.4$, and $200$~GeV.
The particlization hypersurfaces, which at all collision energies correspond to the switching energy density of $\epsilon_{\rm sw} = 0.26$~GeV/fm$^3$, are available via~\cite{MUSICinput}.
For reference, the energy density $\epsilon_{\rm sw} = 0.26$~GeV/fm$^3$ corresponds to the particlization temperature $T_{\rm sw} = 150.6$~MeV at $\mu_B = 0$. In Appendix~\ref{app:esw} other values of $\epsilon_{\rm sw}$ are explored for $\sNN = 200$~GeV which shows that the results exhibit mild dependence on $\epsilon_{\rm sw}$.
The net proton, proton, and antiproton cumulants are calculated in the relevant momentum acceptances analytically, following the method presented in the previous section.
We perform separate calculations employing EV-HRG and ideal HRG models, and study the behavior of cumulants with and without the correction for baryon number conservation.
These different configurations allow us to establish the relevance of repulsive interactions and global baryon conservation.
In Appendix~\ref{app:MC} we perform a cross-check of the analytic results for the case of the ideal HRG model by means of Monte Carlo sampling of hadrons at particlization.

\subsection{Rapidity distributions}

First we look at the net proton rapidity distributions.
To calculate the net proton $dN/dy$ we partition the rapidity axis into bins.
The rapidity density in a given bin then corresponds to the first net proton cumulant evaluated for that bin.
As $dN/dy$ is determined by the mean numbers of particles, it is unaffected by the correction for global baryon conservation.
We observe that net proton rapidity distributions calculated in the EV-HRG and ideal HRG models virtually coincide.
This is attributed to the fact that we match the net baryon density at particlization to the MUSIC output in both scenarios, which leads to virtually identical mean numbers of net protons.

\begin{figure}[t]
  \centering
  \includegraphics[width=.49\textwidth]{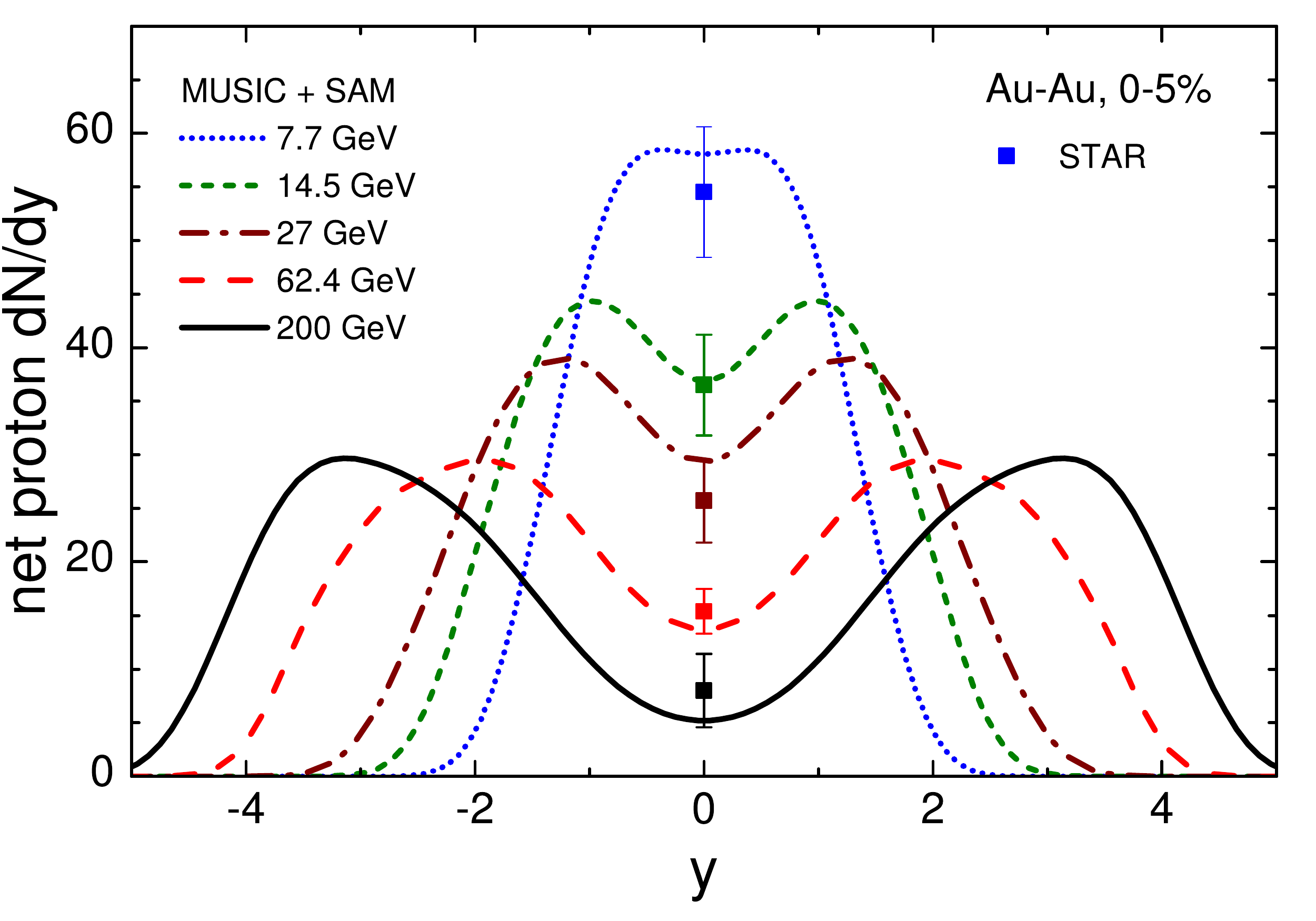}
  \caption{
  Net proton rapidity distributions in 0-5\% Au-Au collisions at various RHIC beam energy scan collision energies.
  The experimental data of the STAR Collaboration~\cite{Abelev:2008ab,Adamczyk:2017iwn,Adam:2019dkq} are shown by the symbols.
  }
  \label{fig:dNdy}
\end{figure}

The resulting rapidity distributions are depicted in Fig.~\ref{fig:dNdy} for various collision energies.
The results agree qualitatively with earlier MUSIC calculations in Ref.~\cite{Shen:2020jwv}.
The calculations also agree within errors with the midrapidity net proton yields measured by the STAR Collaboration~\cite{Abelev:2008ab,Adamczyk:2017iwn,Adam:2019dkq}.
The rapidity dependence of the net proton distribution agrees qualitatively with the experimental data of the BRAHMS Collaboration for $\sNN = 62.4$ and $200$~GeV~\cite{Bearden:2003hx,Arsene:2009aa}, although the data for $\sNN = 200$~GeV is notably overestimated by the model.
The results are also qualitatively consistent with the measurements of the NA49 Collaboration for Pb-Pb collisions at $\sNN = 8.8$ and $17.3$~GeV~\cite{Anticic:2010mp}.

The net proton rapidity distributions show peaks in the forward-backward rapidity regions at all collision energies except for 7.7~GeV.
This reflects the fact that large rapidities probe baryon-rich matter.
It is observed for instance, that larger longitudinal space-time rapidities are characterized by larger values of the baryochemical potential $\mu_B$ at particlization.
This underlines the fact that it is impossible to characterize the whole fireball by a single pair of temperature $T$ and baryon chemical potential $\mu_B$ but instead one has to integrate over different $\mu_B$-$T$ values encompassing the hypersurface.

\subsection{Net proton cumulants}

\begin{figure*}[t]
  \centering
  \includegraphics[width=.49\textwidth]{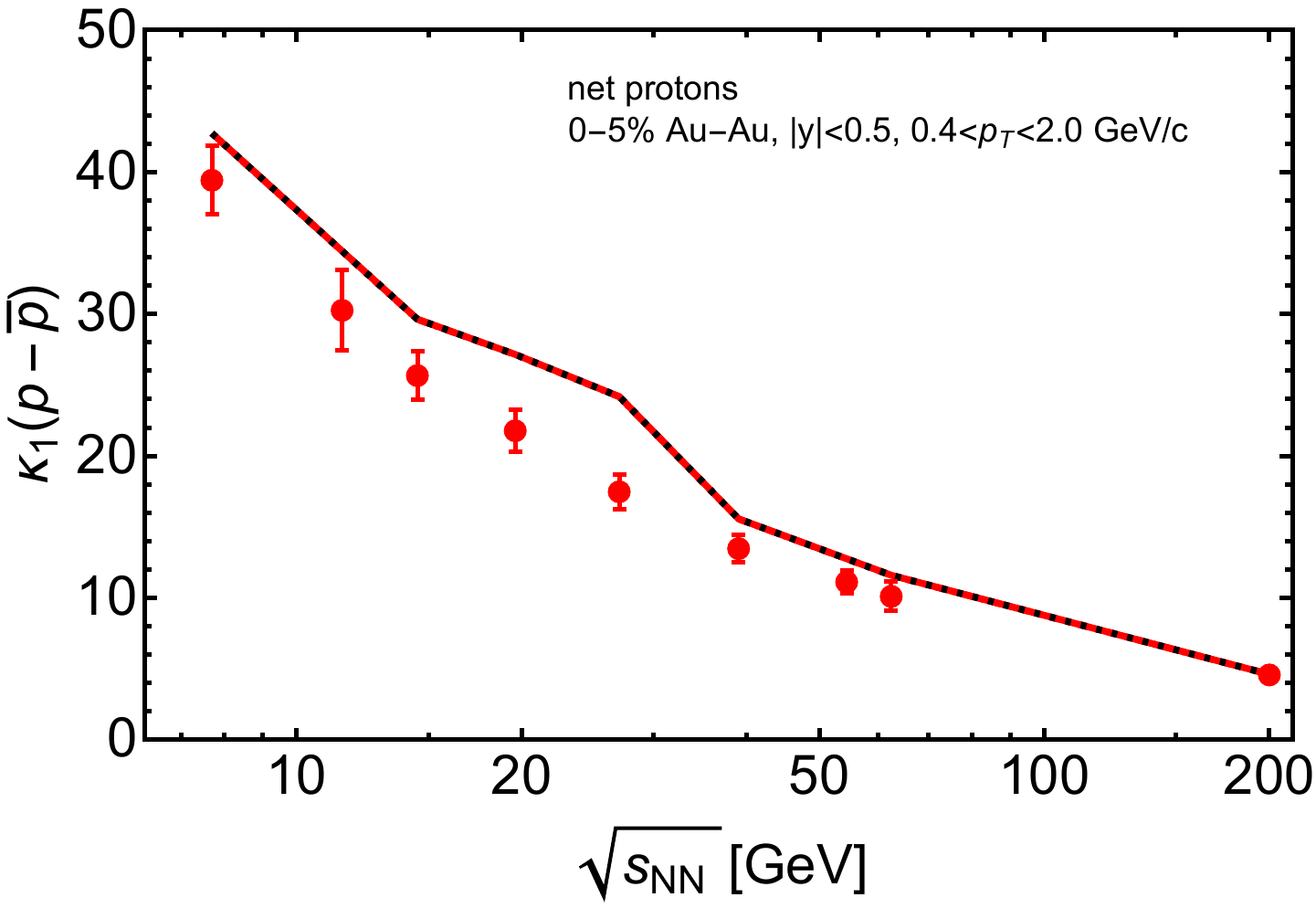}
  \includegraphics[width=.49\textwidth]{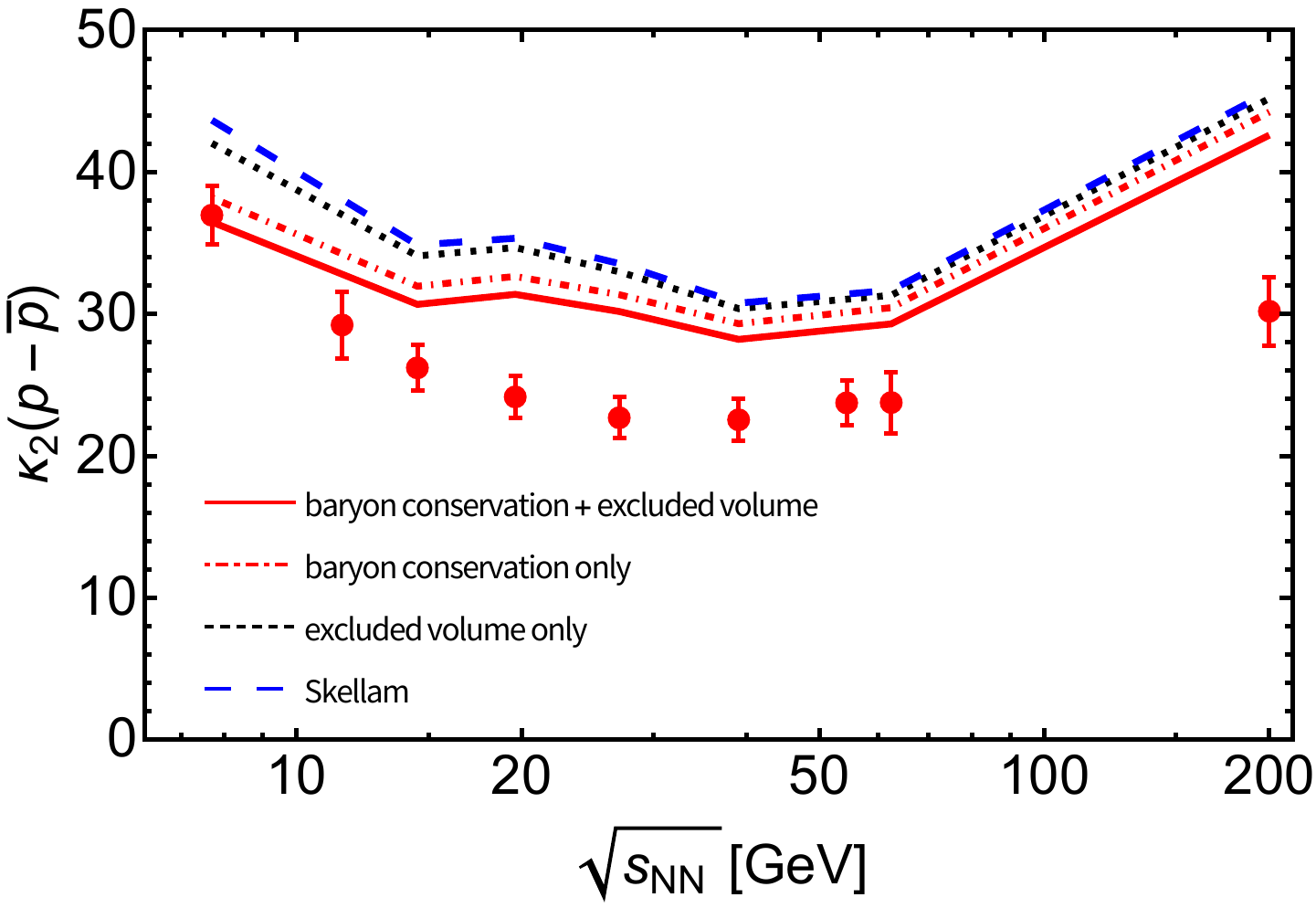}
  \vskip5pt
  \includegraphics[width=.49\textwidth]{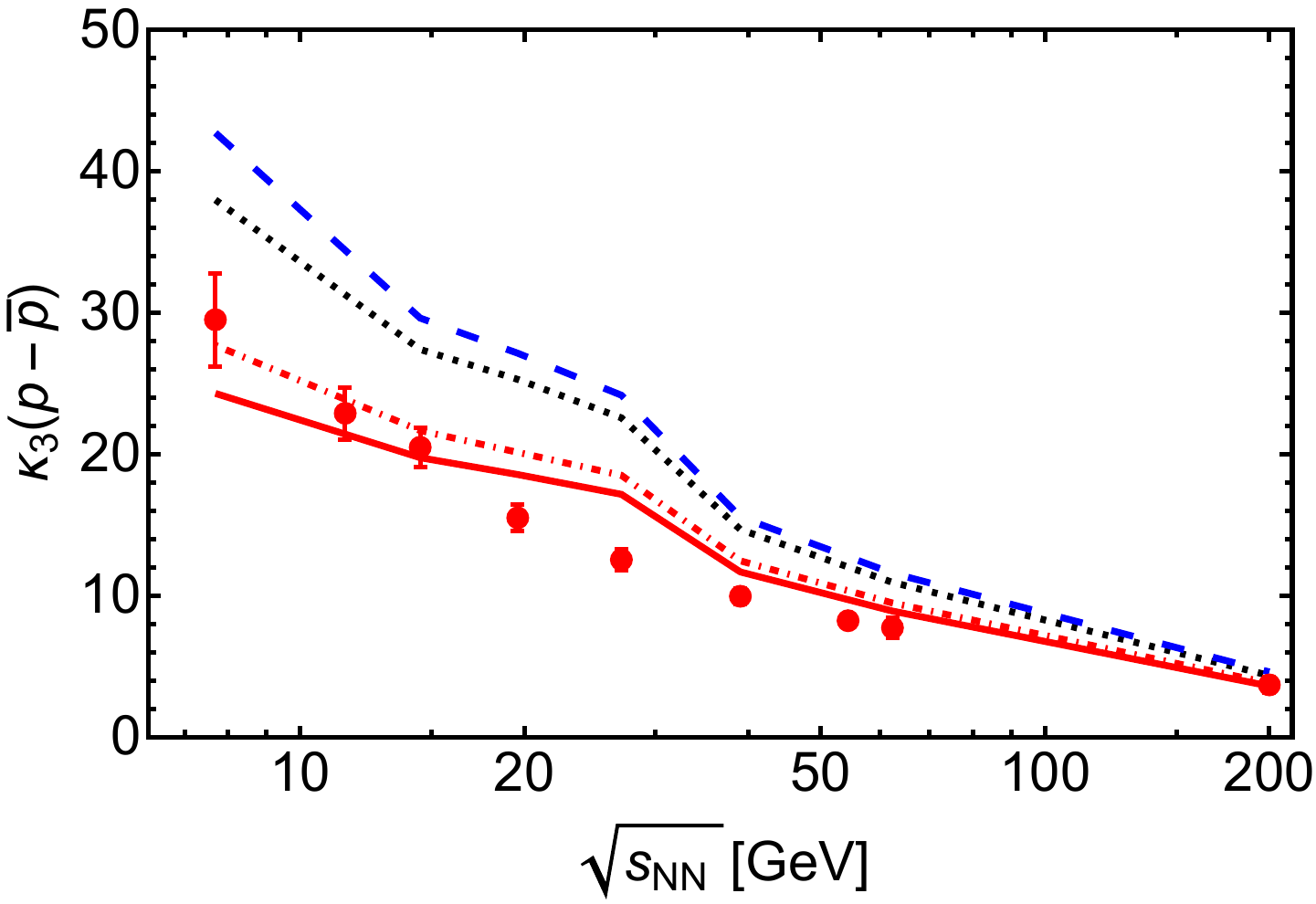}
  \includegraphics[width=.49\textwidth]{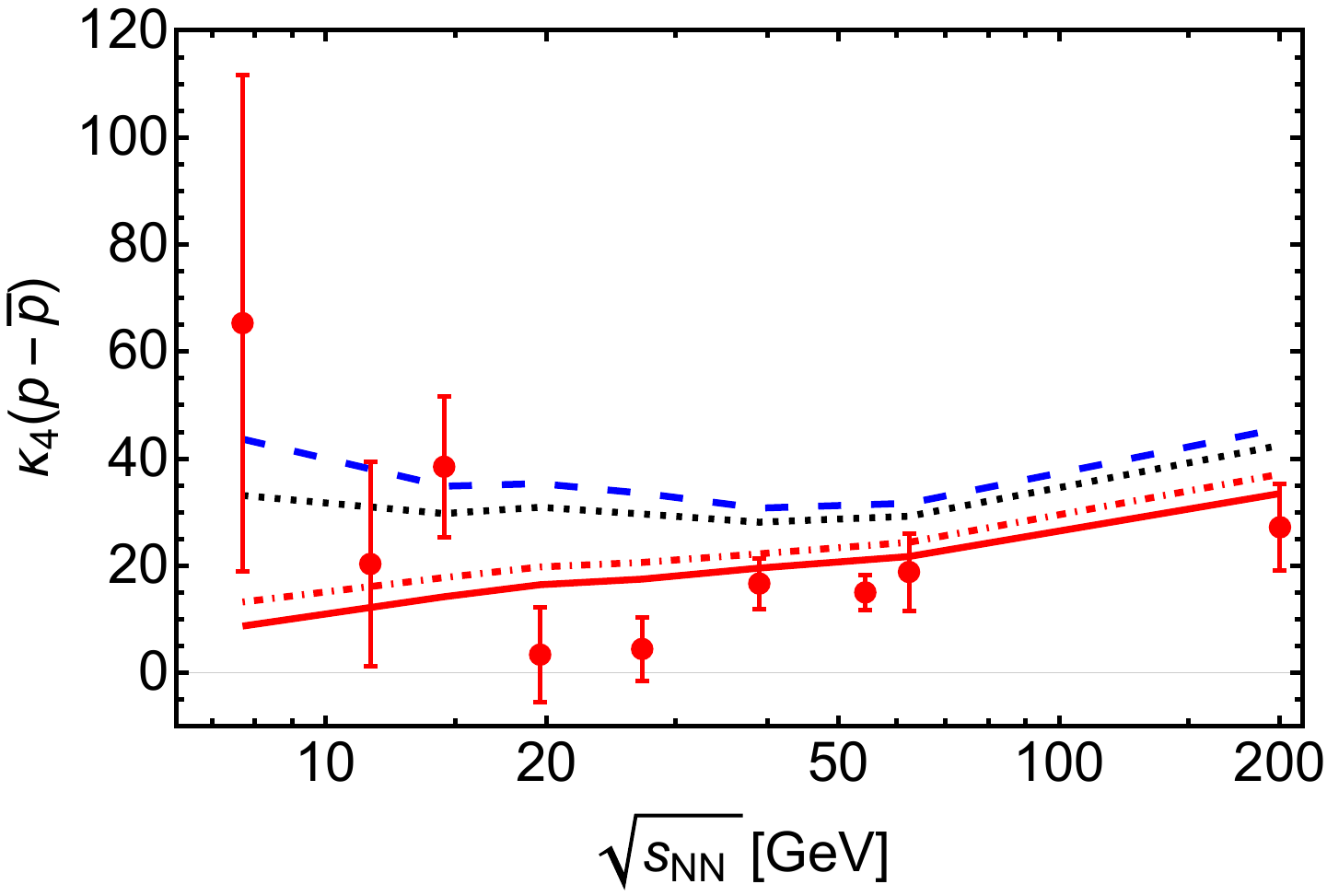}
  \caption{
  Collision energy dependence of the leading four net proton cumulants in 0-5\% Au-Au collisions.
  The calculations are performed using EV-HRG with (solid red lines) and without (dotted black lines) the effect of exact baryon number conservation.
  The dash-dotted red lines correspond to calculations including exact baryon number conservation but neglecting the excluded volume. 
  The dashed lines correspond to the Skellam distribution baseline, i.e. $\kappa_1[p-\bar{p}] = \kappa_3[p-\bar{p}] = \mean{N_p - N_{\bar{p}}}$ and $\kappa_2[p-\bar{p}] = \kappa_4[p-\bar{p}] = \mean{N_p + N_{\bar{p}}}$.
  The experimental data of the STAR Collaboration~\cite{Adam:2020unf} are shown by the red symbols with error bars.
  }
  \label{fig:cumulants}
\end{figure*}

The leading four cumulants of net proton distribution have been measured and presented by the STAR Collaboration in Ref.~\cite{Adam:2020unf} as a function of collision energy.
The measurements were performed in the momentum acceptance $|y| < 0.5$ and $0.4 < p_T < 2.0$~GeV/$c$.
Here we calculate these cumulants in the same momentum acceptance.
The results are presented in Fig.~\ref{fig:cumulants}.  
The calculations are performed within the EV-HRG model with (solid red lines) and without (dotted black lines) the effect of exact baryon number conservation.
We also perform a calculation within the ideal HRG model but including the effect of baryon number conservation~(dash-dotted red line).
The dashed blue lines correspond to the uncorrelated (anti)protons baseline, which is given by the Skellam distribution, i.e. $\kappa_1[p-\bar{p}] = \kappa_3[p-\bar{p}] = \mean{N_p - N_{\bar{p}}}$ and $\kappa_2[p-\bar{p}] = \kappa_4[p-\bar{p}] = \mean{N_p + N_{\bar{p}}}$.

The first net-proton cumulant is unaffected by the correction for global baryon conservation.
It is also virtually unaffected by the excluded volume effects due to the matching of the EV-HRG model equation of state to the net baryon density at particlization.
The model provides a reasonable description of the experimental data, with the possible exception of the 19.6 and 27~GeV points.
The description of these data points can be improved by fine-tuning the simulation parameters.

The second, third, and fourth cumulants are affected by both the excluded volume and baryon conservation, the latter effect being the stronger of the two.
Both effects suppress the cumulants, and the suppression is stronger for higher-order cumulants and lower collision energies.
The effect of excluded volume is stronger at lower collision energies because they probe baryon-rich matter with smaller inter-particle distances between baryons at particlization.
The baryon conservation plays a larger role at smaller energies because a larger fraction of the full baryon number ends up in the midrapidity region which is where the measurements are performed.
Compared to the STAR data, the calculation with simulatenous excluded volume and baryon conservation effects generally yields the best agreement.
The agreement is not perfect everywhere, in particular $\kappa_2[p-\bar{p}]$ is notably overestimated by the model at $\sNN \geq 19.6$~GeV.
This is a reflection of an overestimated mean numbers of protons and antiprotons produced by the model compared to the data.
There are different possible reasons for this.
For instance, if weak decay contributions are overestimated in the model calculation, this may explain the discrepancy.
Although it has been argued that the integrated yields of (anti)protons measured by STAR contain essentially all weak decay contributions~\cite{Oliinychenko:2020znl}, the situation might be different in the measurements of fluctuations.
We performed calculations of the cumulants neglecting all weak decay contributions, and indeed obtain an improved description of the data at some of the energies, although in this case the data are generally underestimated by the model. We did observe, however, that weak decays affect only mildly the various volume-independent ratios of cumulants, thus we keep all weak decay contributions in all our further results throughout this work.
Other potential reasons that may contribute to the overestimation of (anti)proton yields include
neglecting baryon annihilation in the hadronic phase~\cite{Becattini:2012xb}, or a reflection of the so-called thermal proton yield anomaly in the HRG model~\cite{Vovchenko:2018fmh,Andronic:2018qqt}.
As all these possible mechanisms are not linked directly to the QCD equation of state, we shall not study them in detail here but instead look at observables where the effect of describing the total (anti)proton yield is minimized.

\begin{figure*}[t]
  \centering
  \includegraphics[width=.49\textwidth]{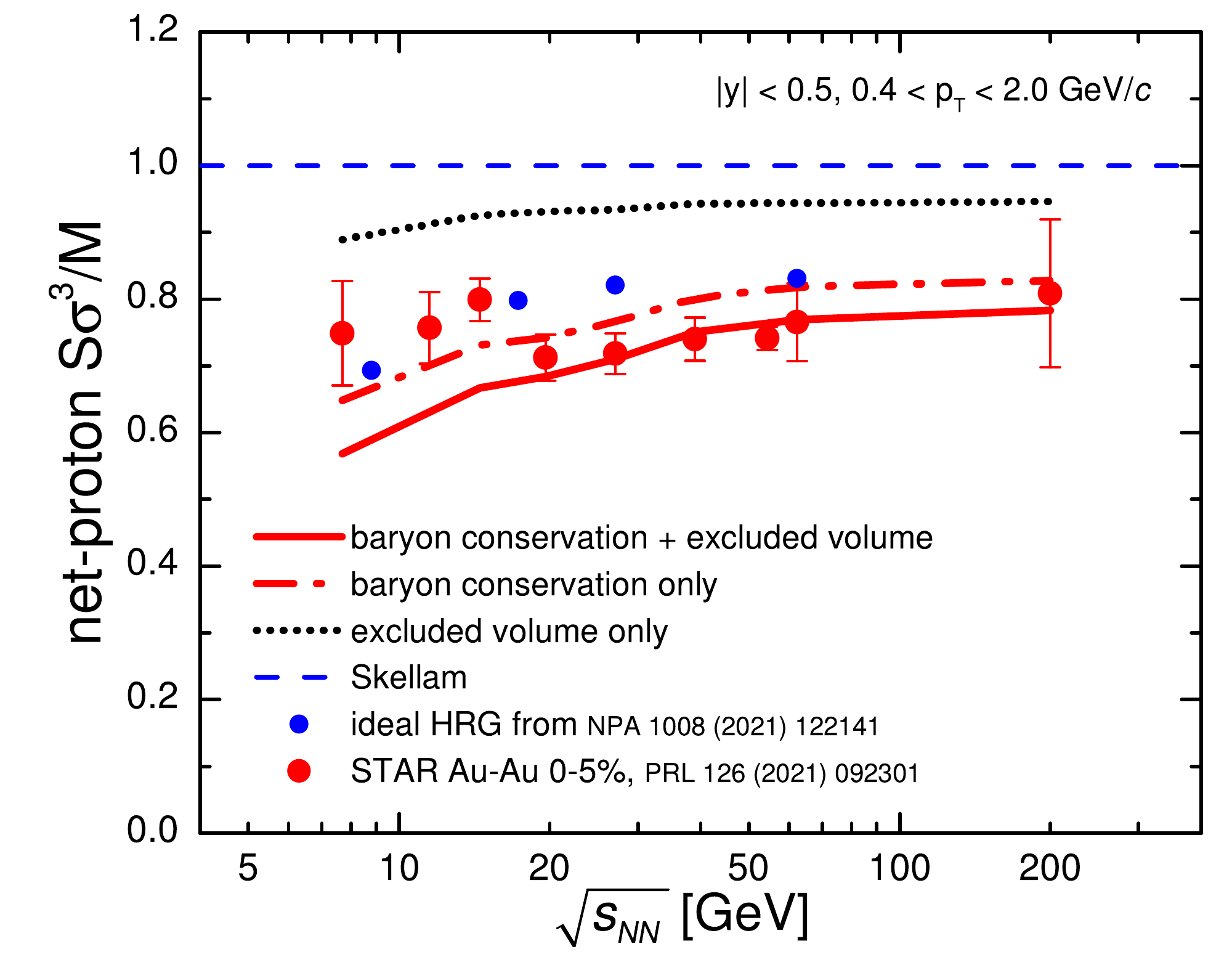}
  \includegraphics[width=.49\textwidth]{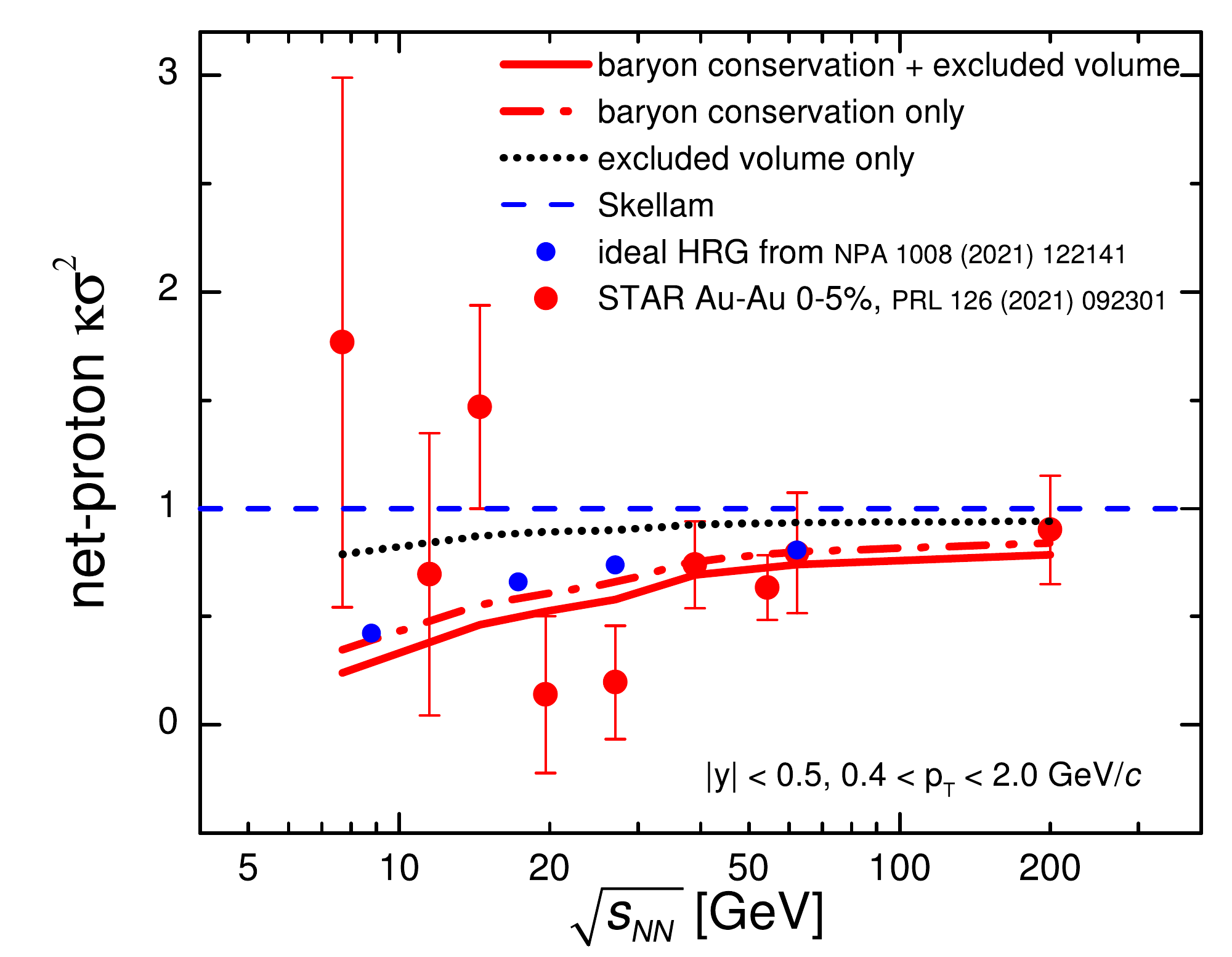}
  \caption{
  Collision energy dependence of the net-proton cumulant ratios $\kappa_3[p-\bar{p}] / \kappa_1[p-\bar{p}] \equiv S\sigma^3 / M$~(left) and $\kappa_4[p-\bar{p}] / \kappa_2[p-\bar{p}] \equiv \kappa \sigma^2$~(right) in 0-5\% Au-Au collisions.
  The red lines depict calculations with~(solid) and without~(dash-dotted) the excluded volume effect, both calculations incorporate the effect of exact baryon conservation. 
  The dotted black lines correspond to calculations incorporating the excluded volume effect, but not exact baryon conservation. 
  The dashed blue lines correspond to the Skellam distribution baselines of unity.
  The experimental data of the STAR Collaboration~\cite{Adam:2020unf} are depicted by the red circles.
  The blue circles correspond to the canonical ensemble ideal HRG model of Ref.~\cite{Braun-Munzinger:2020jbk}.
  }
  \label{fig:cumulantratios}
\end{figure*}

We thus analyze the following ratios of cumulants
\eq{
\frac{S\sigma^3}{M} \equiv \frac{\kappa_3}{\kappa_1}, \qquad \kappa \, \sigma^2 \equiv \frac{\kappa_4}{\kappa_2}~.
}
These two ratios have a baseline of unity in the (Skellam) limit of an uncorrelated production of protons and antiprotons, at any collision energy.
Deviations from unity can be a signal of new physics, in particular, the QCD critical point has been predicted to have a strong influence on these non-Gaussian measures of net proton number fluctuations~\cite{Stephanov:2008qz,Stephanov:2011pb,Mroczek:2020rpm}.
In this sense, $S\sigma^3 / M$ is more convenient than the commonly adopted normalized skewness $s \sigma^2 \equiv \kappa_3 / \kappa_2$ which shows strong collision energy dependence even in the Skellam limit.
Note that the absolute yields of (anti)protons drop out in the ratios of cumulants, thus the ratios are not very sensitive to possible inaccuracies in the description of the overall yields discussed above.
The ratios, however, are sensitive to both the excluded volume and baryon number conservation.

\begin{figure}[t]
  \centering
  \includegraphics[width=.49\textwidth]{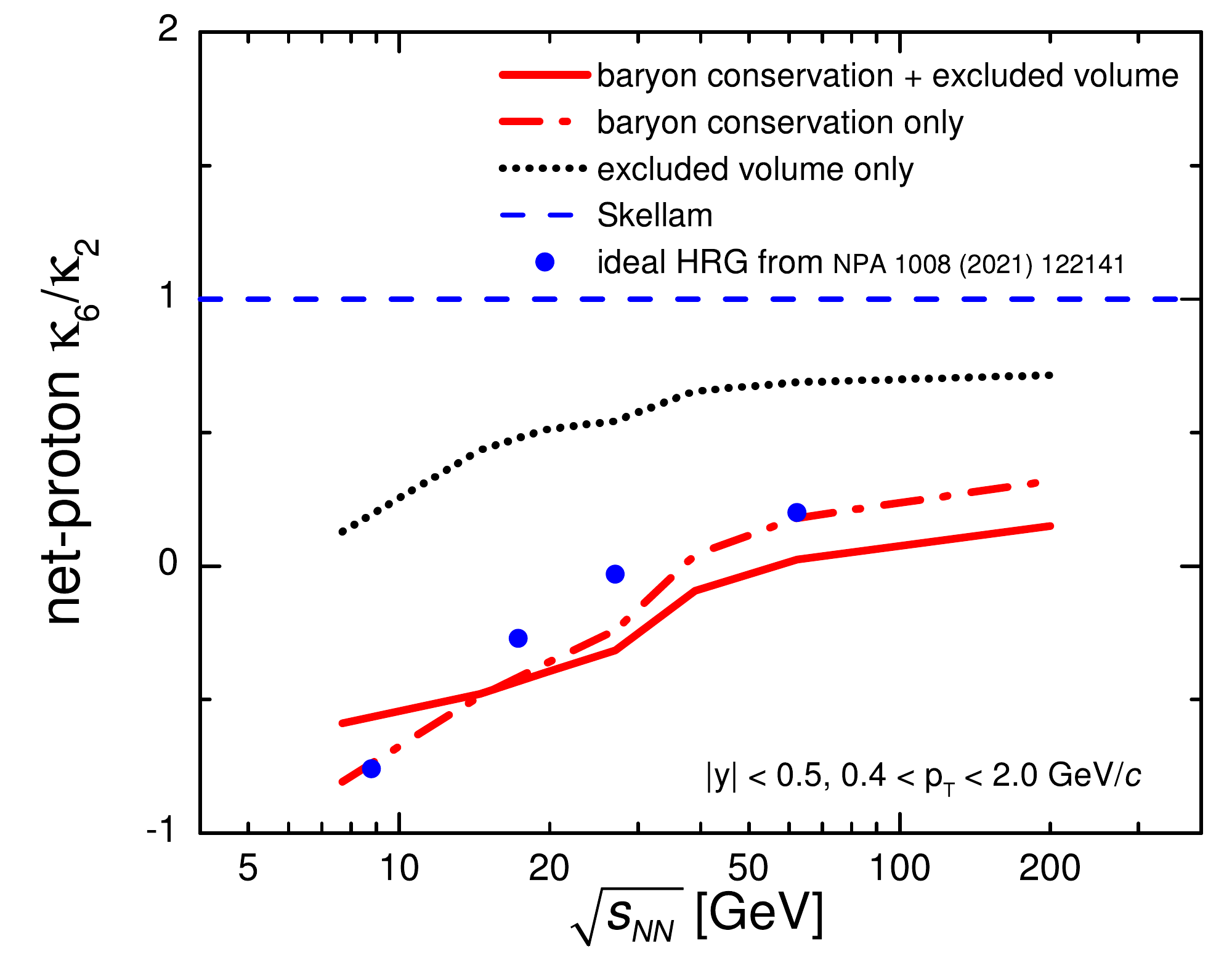}
  \caption{
  Same as Fig.~\ref{fig:cumulantratios} but for the net proton hyperkurtosis $\kappa_6[p-\bar{p}] / \kappa_2[p-\bar{p}]$.
  }
  \label{fig:c6c2}
\end{figure}

Figure~\ref{fig:cumulantratios} depicts the collision energy dependence of $S\sigma^3 / M$ and $\kappa \sigma^2$ calculated in the model along with the experimental data of the STAR Collaboration from Ref.~\cite{Adam:2020unf}.
Both the data and the model calculations show a suppression of $S\sigma^3 / M$ with respect to the Skellam baseline of unity.
When baryon excluded volume, but not global baryon conservation, is incorporated~(dotted black lines), this leads to an improved agreement with the data compared to Skellam, although the suppression of $S\sigma^3 / M$ due to the excluded volume is not sufficient to obtain a quantitative agreement.
Calculations that incorporate global baryon conservation but not excluded volume~(dash-dotted red lines) indicate that the former effect is stronger than the latter one.
In this case the data at $\sNN \lesssim 20$~GeV are described but not at higher collision energies.
Finally, when both the baryon conservation and excluded volume are incorporated, the experimental data at $\sNN \gtrsim 20$~GeV are described on a quantitative level. 
On the other hand, the data at lower collision energies are underestimated.
It should be noted that the magnitude of the excluded volume effects in the EV-HRG model that we use has been constrained using lattice QCD data at $\mu_B = 0$ in Ref.~\cite{Vovchenko:2017xad}.
Thus, we expect the model to be most reliable at the highest collision energies that probe the QCD phase diagram close to the vanishing net baryon density.
Deviations from the data at $\sNN \lesssim 20$~GeV may be an indication of a breakdown of the EV-HRG model that we use. We explore this issue further in the next subsection by studying proton correlation functions.
The behavior of the net proton kurtosis $\kappa \sigma^2$ is qualitatively similar to $S\sigma^3 / M$, although the error bars are considerably larger, especially at the lower collision energies. 
This precludes making strong conclusions from the available data on $\kappa \sigma^2$ from RHIC-BES-I, it should however be possible to use the  more precise future data from RHIC-BES-II for this purpose.

Figure~\ref{fig:c6c2} depicts our predictions for the net proton hyperkurtosis, $\kappa_6/\kappa_2$.
This quantity is strongly suppressed by both the excluded volume and baryon conservation, and it turns negative as the collision energy is decreased to below $\sNN \lesssim 40-60$~GeV.
These predictions can be probed by future high-statistics measurements at RHIC.

We also compare our results with a non-critical baseline of Ref.~\cite{Braun-Munzinger:2020jbk}, which is based on the ideal HRG model and the empirical rapidity distributions.
These results, shown in Figs.~\ref{fig:cumulantratios} and~\ref{fig:c6c2} by the blue points, agree fairly well with our ideal HRG model results with exact baryon conservation~(dash-dotted red lines).
Thus, there is a consistency between our ideal HRG model based calculations and the prior literature.

\subsection{Cumulants versus correlation functions}

More information can be obtained by analyzing cumulants of proton and antiproton distributions separately.
In particular, one can construct the so-called correlation functions~(factorial cumulants) $\hat{C}_n$ from the ordinary cumulants $\kappa_n$~\cite{Bzdak:2016sxg}.
The correlation functions characterize genuine multi-particle correlations. 
For $n > 1$ they vanish in the limit of Poisson statistics (uncorrelated particle production).
The correlation functions are linear combinations of the cumulants:
\eq{
\hat{C}_1 & = \kappa_1, \\
\hat{C}_2 & = -\kappa_1 + \kappa_2, \\
\hat{C}_3 & = 2 \kappa_1 - 3 \kappa_2 + \kappa_3, \\
\hat{C}_4 & = -6 \kappa_1 + 11 \kappa_2 - 6 \kappa_3 + \kappa_4.
}

\begin{figure*}[t]
  \centering
  \includegraphics[width=.69\textwidth]{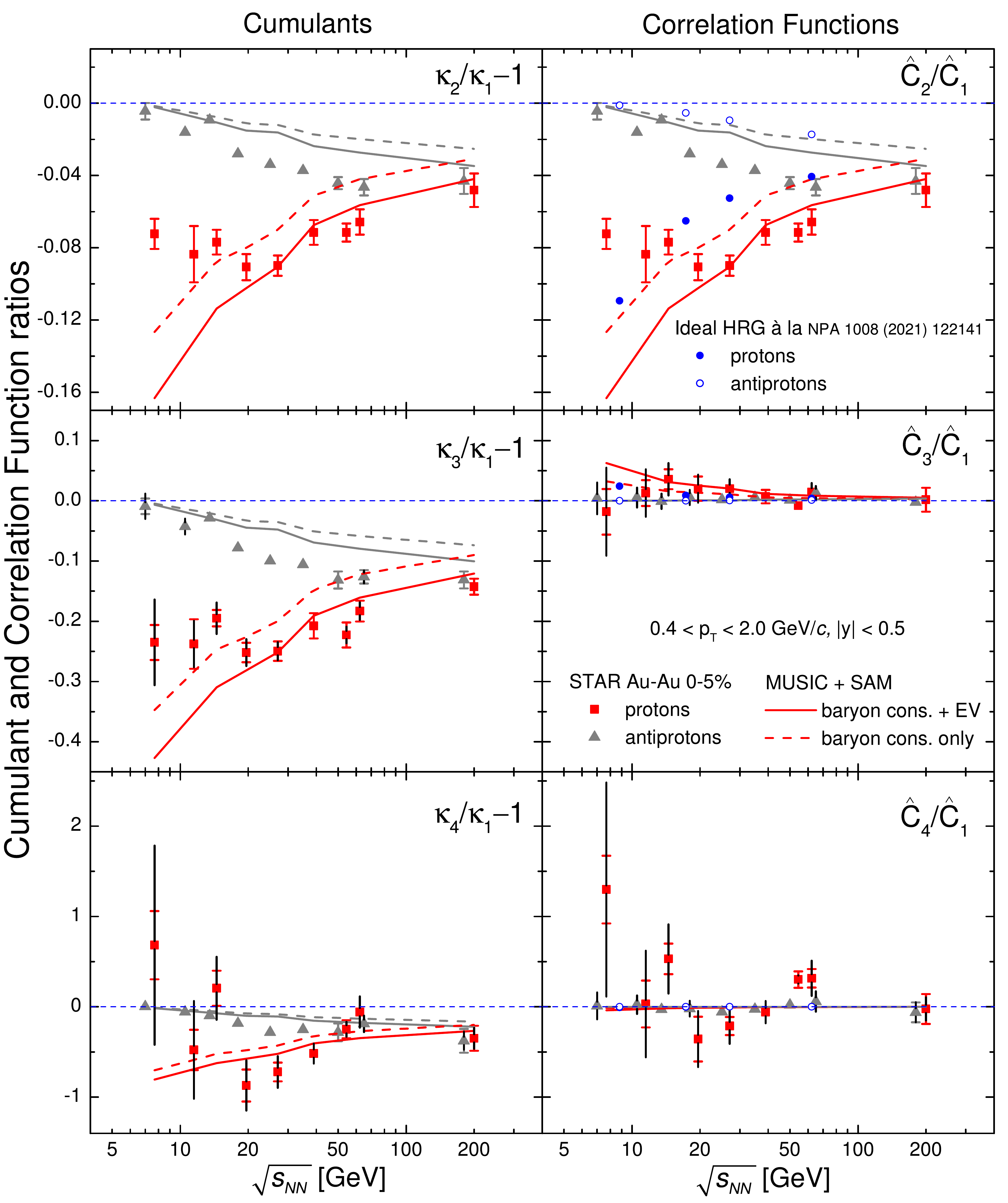}
  \caption{
  Collision energy dependence of scaled (anti)proton cumulants and factorial cumulants~(correlation functions) in 0-5\% Au-Au collisions up to fourth order.
  The solid lines depict calculations that incorporate both the baryon conservation and excluded volume effects~(EV-HRG model) while the dashed lines correspond to baryon conservation only~(ideal HRG model).
  The red squares and gray triangles correspond to the experimental data of the STAR Collaboration~\cite{Abdallah:2021fzj} for protons and antiprotons, respectively.
  The blue circles correspond to the canonical ensemble ideal HRG model calculation based on (anti)proton acceptance fractions from Ref.~\cite{Braun-Munzinger:2020jbk}.
  }
  \label{fig:KandC}
\end{figure*}

Experimental measurements of both the cumulants $\kappa_n$ and the correlation functions $\hat{C}_n$ have recently been presented by the STAR Collaboration in Ref.~\cite{Abdallah:2021fzj}.
Figure~\ref{fig:KandC} depicts the comparison of our model calculations with the experimental data, in terms of normalized quantities, $\kappa_n/\kappa_1 - 1$ and $\hat{C}_n/\hat{C}_1$\footnote{Note that here we follow the notation of Ref.~\cite{Bzdak:2019pkr} and designate cumulants and correlation functions by $\kappa_n$ and $\hat{C}_n$, respectively. This is different from STAR's Ref.~\cite{Abdallah:2021fzj} where this notation is reversed.}.
Deviations of these quantities from zero signal physics beyond the uncorrelated gas of hadrons.

The normalized second order cumulants and correlation functions are equivalent and characterize the two-particle correlations.
The experimental data clearly establish the existence of negative two-particle correlations, both among the protons and the antiprotons.
The data for protons at $\sqrt{s_{\rm NN}} \gtrsim 20$~GeV are adequately described when both the baryon conservation and excluded volume effects are taken into account.
The baryon conservation exerts a stronger influence on $\hat{C}_2/\hat{C}_1$ than the excluded volume although both effects are necessary to obtain a fair agreement with the data at $\sqrt{s_{\rm NN}} \gtrsim 20$~GeV.

At lower energies the suppression of $\hat{C}_2/\hat{C}_1$ is overestimated by the model, especially at $\sNN = 7.7$~GeV.
This can be due to different mechanisms.
For instance, we have neglected the effect of volume fluctuations which would increase $\hat{C}_2/\hat{C}_1$~\cite{Bzdak:2016jxo}.
The STAR Collaboration has applied the centrality bin width correction~\cite{Luo:2013bmi} to minimize the effects of volume fluctuations in the data, which, however, does not remove volume fluctuations completely~\cite{Braun-Munzinger:2016yjz}.
To leading order, the volume fluctuations lead to an additive correction to $\hat{C}_2/\hat{C}_1$~\cite{Skokov:2012ds}, namely
\eq{
\frac{\hat{C}_2^{\rm volF}}{\hat{C}_1^{\rm volF}} = \frac{\hat{C}_2}{\hat{C}_1} + \hat{C}_1 \, v_2. 
}
Here $v_2$ is a normalized variance of volume fluctuations.
The 7.7~GeV STAR data point could be described with volume fluctuations for $v_2 \approx 0.002$, however one would require $v_2 < 0.001$ to not spoil the agreement at higher collision energies.
Thus, the volume fluctuations could only explain the deviations from experimental data if their effect is considerably larger at $\sNN = 7.7$~GeV than at higher energies.
Qualitatively, such a behavior has been indicated before~\cite{Li:2017via} and it remains to be seen if it can provide a quantitative resolution.

A potentially more intriguing explanation for the disagreement at $\sNN = 7.7$~GeV is a presence of sizable attractive interactions among protons. 
This effect is not included in our model and would lead to an increase in $\hat{C}_2/\hat{C}_1$.
If this is the case, the data would indicate a transition from repulsive to attractive proton interaction regime as the collision energy is decreased below $\sNN \simeq 20$~GeV.
One possible mechanism for this would be approaching the QCD critical point.
It should be noted however that approaching the QCD critical point would be also expected to generate multi-particle correlations~\cite{Ling:2015yau}, which has not yet been established by the data.
At lower collision energies one can also expect sizable effects due to the nuclear liquid-gas  transition~\cite{Mukherjee:2016nhb,Vovchenko:2017ayq,Poberezhnyuk:2019pxs}.

It should be noted that at $\sNN = 7.7$~GeV the proton cumulants are expected to also be affected by exact conservation of electric charge.
We estimate this effect in Appendix~\ref{app:BQS} and show that this would lead to a further suppression of $\hat{C}_2/\hat{C}_1$ by about 20\%.
This would increase the disagreement with the experimental data.

The trends in the data for antiprotons are reproduced by the model, although, in contrast to the protons, the data point to considerably stronger anticorrelation among the antiprotons than predicted by the model.
This difference between the protons and antiprotons may be related to their possibly different production mechanisms.
While the measured protons consist of both the stopped and the newly produced protons, all the measured antiprotons are the newly produced particles only.
If the newly produced particles are affected by a different mechanism compared to the stopped protons, for instance by local rather than global baryon conservation, this may lead to a difference in the behavior of two-particle correlations of protons and antiprotons.
A more involved modeling, however, is required to shed light on this possibility.
It should also be noted that the agreement of our present model with the data is already much better than that of the UrQMD model calculations~\cite{Xu:2016qjd,He:2017zpg} shown in the STAR paper~\cite{Abdallah:2021fzj}.
The results based on the non-critical baseline of Ref.~\cite{Braun-Munzinger:2020jbk} are shown in Fig.~\ref{fig:KandC} by the blue points.
They agree well with our ideal HRG model based results and thus show a similar quantitative disagreement with the STAR data for the antiprotons.

The higher order correlation functions $\hat{C}_3/\hat{C}_1$ and $\hat{C}_4/\hat{C}_1$ show only small deviations from zero in our model.
This is consistent with the fact that our model has no critical point and the associated critical fluctuation dynamics which would be expected to generate strong multi-particle correlations among protons in the vicinity of the critical point~\cite{Ling:2015yau}.
The result is also consistent with an earlier observation made in Ref.~\cite{Bzdak:2016jxo} that baryon conservation, which is included in our model, has a modest effect on three- and four-proton correlation functions.
Our results for $\hat{C}_3/\hat{C}_1$ and $\hat{C}_4/\hat{C}_1$ are consistent with the STAR data, if the experimental error bars are to be taken seriously.
This also means that the statistically significant deviations of the third and fourth order cumulants from the Skellam baseline are indeed driven by the two particle correlations, i.e. by the contributions of $\hat{C}_2$ to $\kappa_3$ and $\kappa_4$ rather than by genuine multi-particle correlations. Thus, these data presently show no indication for the existence of the QCD critical point in the studied collision energy regime.
As the experimental uncertainties at $\sNN \lesssim 14.5$~GeV are sizable, however, the present data also do not rule out a possible presence of notable multi-particle correlations among protons in that collision energy regime. 
The high statistics data coming from RHIC-BES-II will thus be able to shed light on this issue.

\subsection{Acceptance dependence}

The cumulants and correlation functions have been measured by the STAR Collaboration as a function of acceptance in rapidity.
Here we compare our model predictions for the acceptance dependence of cumulants with the STAR data.
As neither the model nor the STAR data show conclusive notable deviations from zero for the higher order normalized correlation functions $\hat{C}_3/\hat{C}_1$ and $\hat{C}_4/\hat{C}_1$, we focus the analysis of the acceptance dependence on the second normalized correlation function $\hat{C}_2/\hat{C}_1$.

\begin{figure}[t]
  \centering
  \includegraphics[width=.49\textwidth]{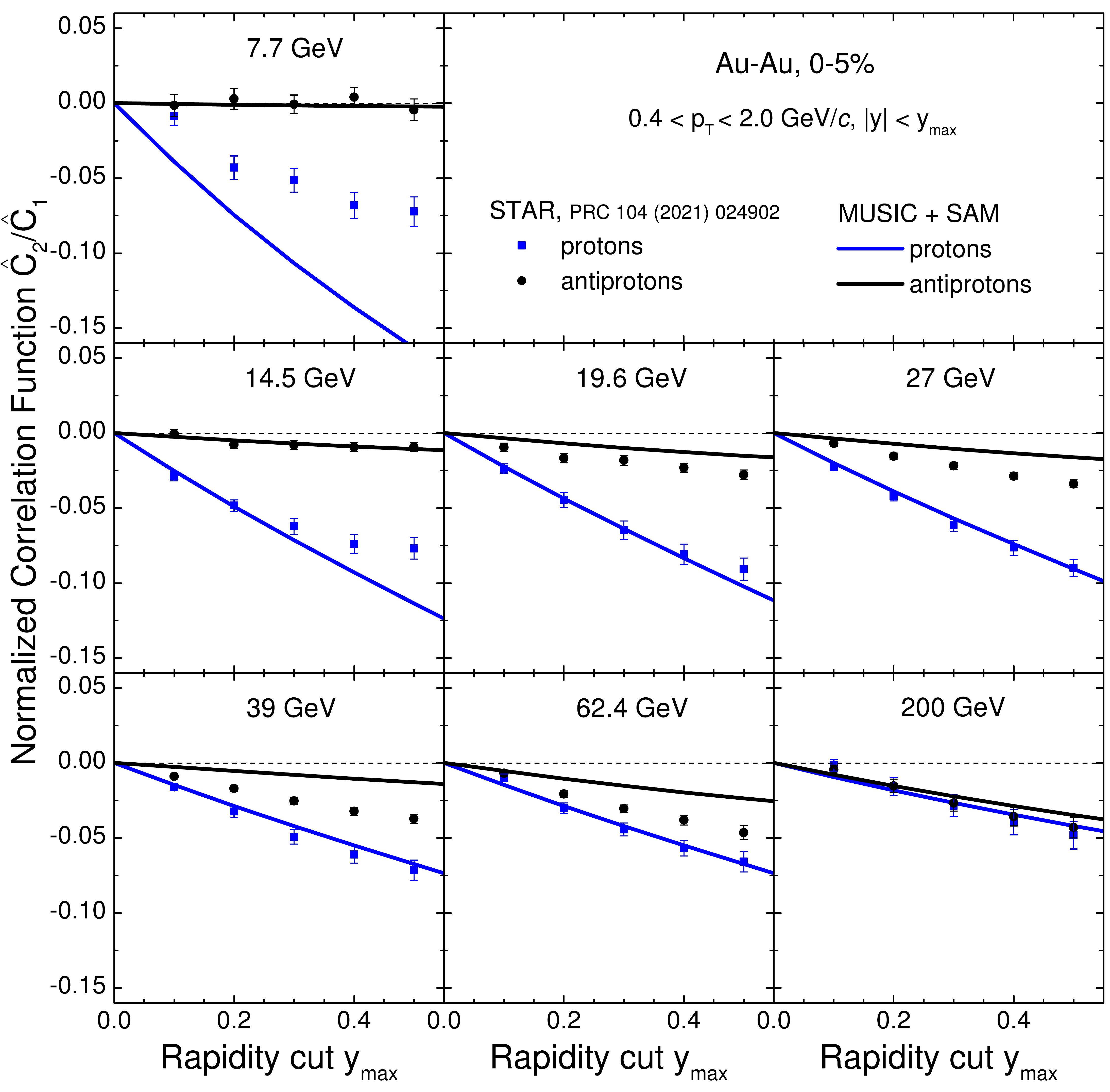}
  \caption{
  Rapidity acceptance dependence of the second normalized factorial cumulant~(correlation function) $\hat{C}_2/\hat{C}_1$ of protons~(blue lines) and antiprotons~(black lines)
  calculated from hydrodynamics and compared to the experimental data of the STAR Collaboration~\cite{Abdallah:2021fzj}.
  The calculations incorporate both the excluded volume effect and global baryon conservation.
  }
  \label{fig:dYacc}
\end{figure}

The results for proton and antiproton number $\hat{C}_2/\hat{C}_1$ as function of the rapidity cut $y_{\rm max}$~(i.e. $|y| < y_{\rm max}$) are shown in Fig.~\ref{fig:dYacc}.
The magnitude of $\hat{C}_2/\hat{C}_1$ increases with $y_{\rm max}$ for both the protons and antiprotons, at all collision energies.
This is the expected result which reflects that (i) the effect of baryon conservation becomes stronger when larger fraction of particles is accepted~\cite{Bzdak:2012an} and (ii) the thermal smearing of local correlations diminishes for larger $y_{\rm max}$~\cite{Ling:2015yau}.

For protons, the STAR data at $\sNN = 19.6$~GeV and above are described by the model quite well, including the  slope of the $y_{\rm max}$ dependence.
For $\sNN = 14.5$~GeV the data are described up to $y_{\rm max} = 0.3$, whereas at $y_{\rm max} = 0.4$ and $0.5$ the model predictions are below the data, i.e. the slope in the data changes faster than in the model.
For $\sNN = 7.7$~GeV the model is below the data for all $y_{\rm max}$, the largest deviations being observed at the maximum measured $y_{\rm max} = 0.5$.
Interestingly, the data at $\sNN = 7.7$ and $14.5$~GeV~(as well as at $\sNN = 11.5$~GeV~\cite{Abdallah:2021fzj} not shown here) show indications that the slope of the $y_{\rm max}$ dependence of proton $\hat{C}_2/\hat{C}_1$ may flip sign at $y_{\rm max} > 0.5$.
Such a qualitative feature is not observed in our model, i.e. it cannot be attributed to baryon conservation or baryon repulsion.

The STAR data for antiprotons are described at the lowest two energies, $\sNN = 7.7$~GeV and $14.5$~GeV, as well as at the top RHIC energy $\sNN = 200$~GeV.
At the intermediate energies the magnitude of the negative two-particle correlation among the antiprotons is underestimated by the model, for all values of $y_{\rm max}$.
The data indicate a larger negative slope of antiproton  $\hat{C}_2/\hat{C}_1$ than predicted by the model at these energies.

\subsection{Centrality dependence}

Our calculations have been focused on 0-5\% Au-Au collisions, as in that regime the assumptions of our formalism are most appropriate. Namely, the degree of thermal and chemical equilibration is expected to the highest in most central collisions, the effect of volume fluctuations smaller than in peripheral collisions, and the applicability conditions of SAM-2.0 satisfied with the highest accuracy.
Nevertheless, for the sake of completeness, we have also performed calculations of proton cumulants at different centralities within our event-averaged hydrodynamics framework.
We find that all cumulant ratios stay essentially flat as function of centrality at a given collision energy, i.e. our framework predicts essentially no centrality dependence for all volume-independent measures of event-by-event fluctuations.
This observation agrees well with the measurements of the STAR Collaboration~\cite{Abdallah:2021fzj} for $\sNN \gtrsim 20$~GeV and $\mean{N_{\rm part}} \gtrsim 100$, while at lower energies and in peripheral collisions deviations from this picture are observed, indicating that a more involved modeling is warranted in those regimes.

\subsection{Protons vs baryons}

Here we discuss an important issue which affects many theory-to-experiment comparisons, namely the difference between baryon and proton number fluctuations.
The experiment has direct access to the latter but it is notoriously difficult to measure all baryons.
On the other hand, proton number is inaccessible in many (effective) QCD theories, for instance lattice QCD.
Cumulants of net baryon number are computed instead, and often directly compared to net proton cumulants measured in the experiment.

\begin{figure}[t]
  \centering
  \includegraphics[width=.49\textwidth]{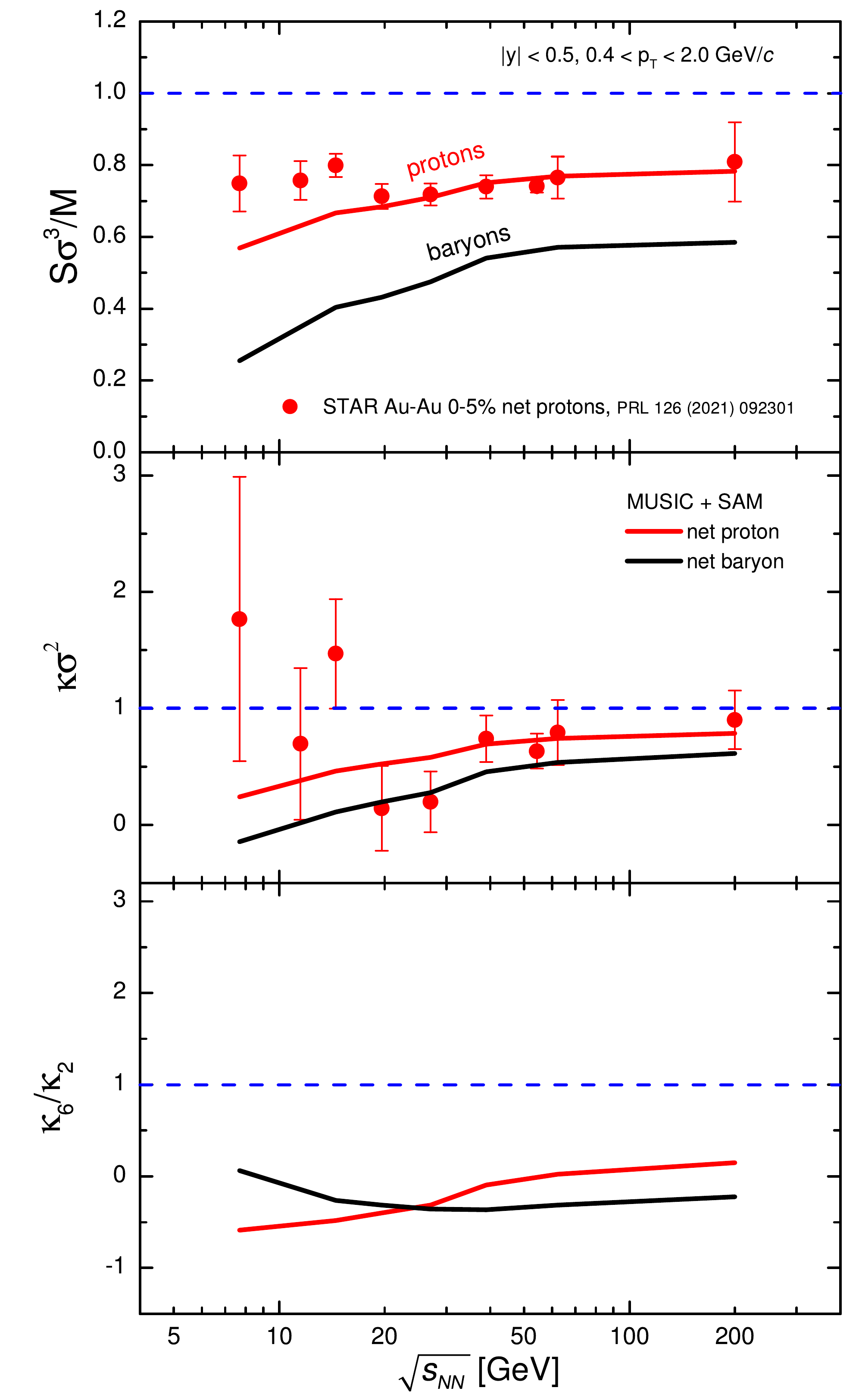}
  \caption{
  Collision energy dependence of
  $S \sigma^3/M$, $\kappa \sigma^2$, and $\kappa_6/\kappa_2$ of net protons~(red lines) and net baryons~(black lines) in 0-5\% central Au-Au collisions.
  The calculations incorporate both the baryon conservation and excluded volume effects.
  The experimental data of the STAR Collaboration~\cite{Adam:2020unf} are depicted by the red circles.
  }
  \label{fig:pvsB}
\end{figure}

In our model it is possible to compute both the proton and baryon number cumulants. 
This then allows us to establish to what extent the two correspond to each other for conditions realized in heavy-ion collisions at RHIC-BES.
Figure~\ref{fig:pvsB} depicts the beam energy dependence of $S \sigma^3/M$, $\kappa \sigma^2$, and $\kappa_6/\kappa_2$ of net protons~(red lines) and net baryons~(black lines) calculated within our formalism including the excluded volume and baryon conservation effects.
It is seen that net proton and net baryon cumulants are quite different, with baryons generally exhibiting larger deviations from the Skellam baseline.
In particular, the net baryon $S \sigma^3/M$ disagrees with the experimental data on net protons at all collision energies, whereas the net proton calculation agrees much better.
The error bars in the data for $\kappa \sigma^2$ are still too large to make a clear distinction between the model predictions for net protons and net baryons.
However, from the difference between red and black lines it is clear that the issue will persist once precision measurements of $\kappa \sigma^2$ become available.

Qualitatively, $S \sigma^3/M$ and $\kappa \sigma^2$ exhibit similar trends as functions of collision energy when compared between protons and baryons.
As seen in the bottom panel of Fig.~\ref{fig:pvsB}, this is no longer the case for $\kappa_6/\kappa_2$, where the net proton hyperkurtosis monotonically increases with $\sNN$ while the net baryon hyperkurtosis exhibits a non-monotonic dependence.
Note that this nonmonotonicity here is unrelated to critical phenomena, which our model does not have, but is caused by an interplay between baryon repulsion and conservation which is sensitive to the collision energy.

\begin{figure}[t]
  \centering
  \includegraphics[width=.49\textwidth]{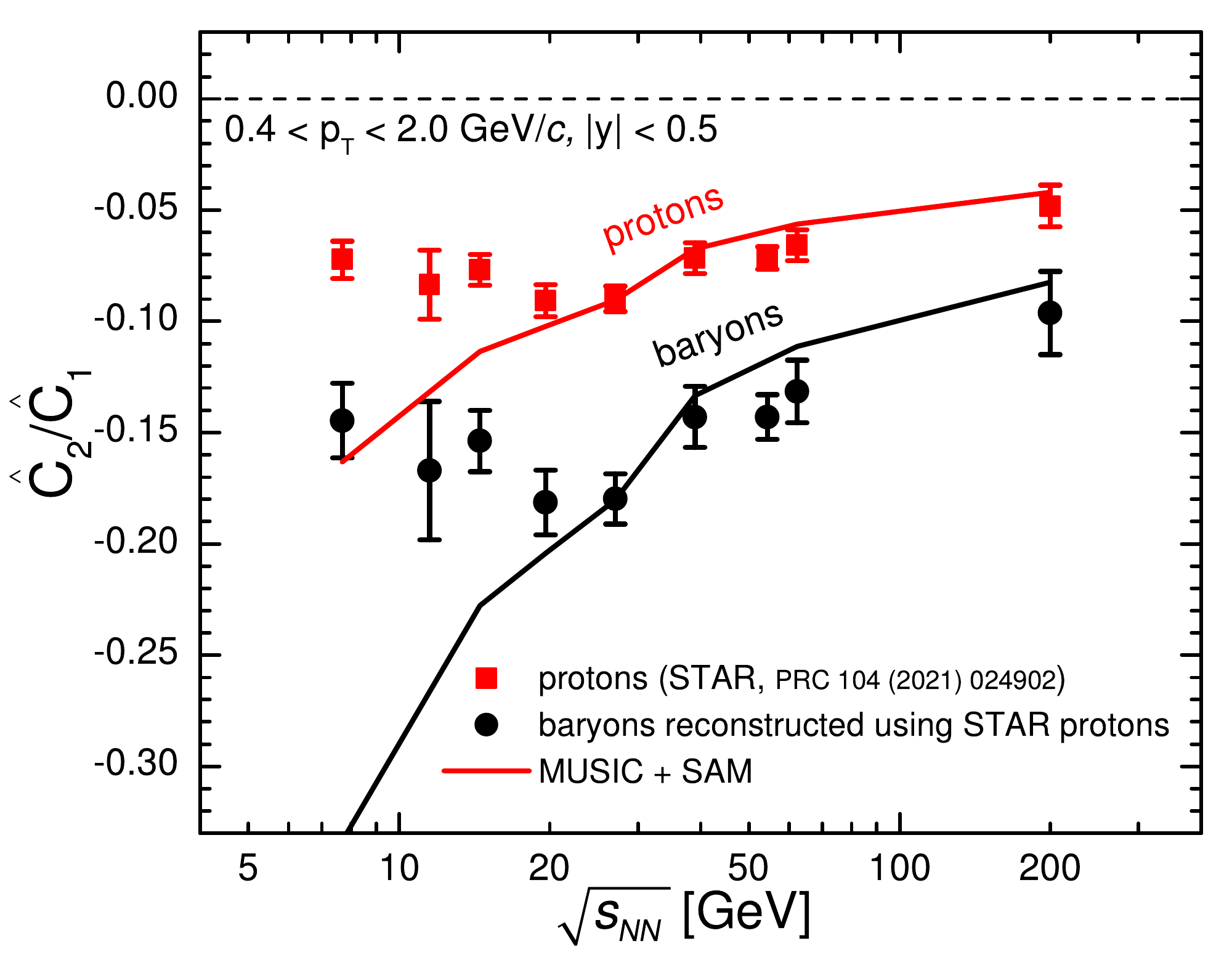}
  \caption{
  Collision energy dependence of the second scaled factorial cumulant
  $\hat{C}_2/\hat{C}_1$ of protons~(red line) and baryons~(black line) in 0-5\% central Au-Au collisions.
  The calculations incorporate both the baryon conservation and excluded volume effects.
  The red squares depict the experimental data of the STAR Collaboration for protons~\cite{Abdallah:2021fzj} while the black circles correspond to baryons reconstructed from the proton data using the binomial distribution.
  }
  \label{fig:C2-p-vs-B}
\end{figure}

Note that a method exists to reconstruct baryon number fluctuations from the measured proton number fluctuations~\cite{Kitazawa:2011wh,Kitazawa:2012at}.
This method assumes that the isospin is randomized at the final stage of the collision, for instance due to $\Delta$ resonance formation and regeneration reactions during the hadronic phase.
In that case baryon number cumulants can be obtained by performing a binomial unfolding of the measured proton cumulants.
This essentially corresponds to an additional binomial efficiency correction with the efficiency probability $q \approx 1/2$, as discussed in Refs.~\cite{Kitazawa:2011wh,Kitazawa:2012at,Bzdak:2012ab}.
Experiment can do this procedure, however this requires very precise measurements of high-order cumulants, because the binomial unfolding increases the error in $\kappa_n$ by factor of order $q^{-n} \approx 2^n$.
Thus applying the unfolding may not be useful using the presently available data on the skewness and kurtosis of (net-)proton distributions which have sizable error bars.

On the other hand, the second order cumulants have already been measured fairly precisely, making the  baryon unfolding procedure feasible to do.
To illustrate this, we apply the unfolding to the STAR data on the ratio $\hat{C}_2/\hat{C}_1$ of proton number factorial cumulants.
Following~\cite{Kitazawa:2011wh,Kitazawa:2012at}, the baryon and proton factorial cumulants are related by $\hat{C}^B_n =  \hat{C}^p_n / q^n$. Thus,
\eq{\label{eq:CBp}
\frac{\hat{C}^B_2}{\hat{C}^B_1} = \frac{1}{q} \, \frac{\hat{C}^p_2}{\hat{C}^p_1}~.
}
We apply Eq.~\eqref{eq:CBp} with $q = 1/2$ to the STAR data to reconstruct two-particle correlations of the baryon number. 
The result is depicted in Fig.~\ref{fig:C2-p-vs-B}.
The baryon number $\hat{C}_2/\hat{C}_1$ constructed from the STAR data agrees well with our model calculation at $\sNN \gtrsim 20$~GeV, similarly to the agreement for the proton number $\hat{C}_2/\hat{C}_1$.

The results presented in this section indicate that there are essential quantitative differences between proton and baryon number cumulant ratios, and that the two may not be compared directly.
Either baryon number cumulants should be reconstructed from the data on proton number cumulants via the method of Refs.~\cite{Kitazawa:2011wh,Kitazawa:2012at}, or the proton number cumulants, rather than baryon number cumulants, should be calculated in the theory.

\section{Discussion and summary}
\label{sec:summary}

In this work we calculated the leading six (net-)(anti)proton cumulants in heavy-ion collisions at RHIC beam energy scan energies in the framework of relativistic viscous hydrodynamics.
The cumulants have been calculated in the momentum acceptance where experimental measurements have been performed by the STAR Collaboration.
For the first time, effects of exact global baryon conservation and repulsive interactions among baryons, modeled by excluded volume, have been incorporated simultaneously and in a dynamical description of heavy-ion collisions.
The excluded volume parameter has been chosen such that deviations from the Skellam distribution in net baryon cumulants computed in lattice QCD at $\mu_B = 0$ are reproduced by the excluded volume HRG model used in our analysis.
The model does not incorporate any critical point effects at finite baryon density, thus, our calculations correspond to the no-critical-point scenario.

We obtain good agreement with the experimental data of the STAR Collaboration on net-proton cumulant ratios $S \sigma^3 / M$ and $\kappa \sigma^2$ at $\sNN \gtrsim 20$~GeV. 
It is observed that the effect of baryon conservation has a stronger influence on the proton cumulants than excluded volume, although incorporating both effects simultaneously is required in order to reproduce the experimental data  on $S \sigma^3 / M$ quantitatively.
At lower collision energies, $\sNN \lesssim 20$~GeV, the data are underestimated by the full model.
The model is in fair agreement with the data on net proton $\kappa \sigma^2$, although the available data has rather large error bars.
Our model predicts that the net proton hyperkurtosis $\kappa_6 / \kappa_2$ in the STAR acceptance $|y| < 0.5$, $0.4 < p_T < 2.0$~GeV/$c$ turns negative at $\sNN \lesssim 40$~GeV, mainly as a consequence of strong effect of baryon conservation.

We explored the behavior of cumulants and factorial cumulants of proton and antiproton distributions.
It is observed that our model produces notable negative two-particle correlations among protons and antiprotons, but only mild three- and four-particle correlations, characterized by small values of the third and fourth scaled factorial cumulants, $\hat{C}_3/\hat{C}_1$ and  $\hat{C}_4/\hat{C}_1$.
In this case the behavior of the high-order cumulants such as skewness and kurtosis is driven by the two-particle correlations rather than by multi-particle correlations that would have been expected near the critical point.
The experimental data are consistent with small, if not vanishing, three- and four-particle proton correlations within errors, thus these data do not contain statistically significant evidence for the existence QCD critical point in the studied collision energy regime.
Note, however, that the error bars on $\hat{C}_3/\hat{C}_1$ and especially $\hat{C}_4/\hat{C}_1$ are quite large, and it is possible that more precise measurements may find evidence for multi-particle correlations.
The upcoming data from RHIC-BES-II will be essential to shed light on this possibility.

The experimental data for the second normalized correlation function $\hat{C}_2/\hat{C}_1$ of protons is described well by the model at energies $\sNN = 19.6$~GeV and above.
At lower energies the model predictions are notably below the data.
We discussed volume fluctuations and/or attractive interactions as a possible explanation for this discrepancy but further studies are required to shed more light on this issue.
It has also been observed that negative two-particle correlations among antiprotons seen in STAR data are notably underestimated by the model at $\sNN = 11.5$~GeV and $19.6$-$62.4$~GeV, whereas at $\sNN = 7.7$, $14.5$, and $200$~GeV the data are described well.
An explanation of this observation is presently an open question.

Compared to other quantitative theoretical predictions for the cumulants that are available in the literature, our model demonstrates a much better agreement with the STAR data than the UrQMD transport model calculations shown in~\cite{Abdallah:2021fzj}.
Our calculations that incorporate baryon conservation but not the excluded volume repulsion~(ideal HRG model) are consistent with the data-driven approach of Ref.~\cite{Braun-Munzinger:2020jbk} within the same ideal HRG model framework.
We do observe, however, that the quantitative agreement with the STAR data on net proton $S\sigma^3 / M$ and proton number normalized correlation function is obtained at $\sNN \gtrsim 20$~GeV only when the baryon repulsion is incorporated in addition to baryon conservation.
The presence of baryon repulsion is in line with the behavior of baryon number susceptibilities at $\mu_B = 0$ observed in lattice QCD.

Comparisons between (net-)proton and (net-)baryon cumulants revealed essential quantitative differences between the two in the RHIC-BES regime.
The higher-order net baryon cumulants generally exhibit larger deviations than the net proton ones from the Skellam distribution baseline of an uncorrelated particle production.
This is due to the fact that protons form a subset of all baryons, thus the strength of the measured correlations is diluted compared to the full baryon set.
As a consequence, for instance, the calculated net baryon $S\sigma^3 / M$ underestimates significantly the measured net proton $S\sigma^3 / M$, whereas the calculated net proton $S\sigma^3 / M$ agrees well with the data.
It is thus essential that the same quantities are employed for theory-to-experiment comparisons, in particular, we find no justification for direct comparisons between the measured (net-)proton and calculated (net-)baryon cumulants that have often been performed in the literature.

One way to address the issue of the difference between proton and baryons cumulants is to unfold the baryon cumulants from the proton ones using the method of Kitazawa and Asakawa~\cite{Kitazawa:2011wh,Kitazawa:2012at}.
In the present work we have demonstrated the feasibility of this procedure by unfolding the scaled factorial cumulant $\hat{C}_2^B / \hat{C}_1^B$ of baryons from the corresponding scaled factorial cumulant $\hat{C}_2^p / \hat{C}_1^p$ of the protons that was measured by the STAR Collaboration.
The resulting data on baryon number $\hat{C}_2^B / \hat{C}_1^B$ agrees reasonably well with our model calculations of this quantity.
The challenge in applying the unfolding to high-order cumulants lies in the fact the this procedure significantly increases the resulting experimental uncertainties.
For this reason the method may not be viable for the presently available data from RHIC-BES phase I but should be viable to do using the more precise data coming from phase II.
We hope that results presented here will serve as a motivation for this procedure to be done.

We have not incorporated any critical fluctuation dynamics associated with the QCD critical point in our study.
In that sense, our results can be viewed as a baseline that incorporates essential non-critical contributions to (net-)proton number cumulants stemming from baryon number conservation and repulsive baryonic interactions.
Unambiguous signals of the QCD critical point in the beam energy scan regime, if there is one to be found, shall manifest themselves as deviations from our model calculations.
We view the three- and four-particle correlation functions (factorial cumulants) of proton number to be particularly promising in that regard, perhaps more so than the ordinary cumulants.
Our model, which has no critical fluctuations, predicts these scaled factorial cumulants to be small.
On the other hand, the  multi-particle correlations among protons are expected to be strong in the vicinity of the critical point~\cite{Ling:2015yau} and may well be reflected in a sizable magnitude of the corresponding scaled factorial cumulants such as $\hat{C}_3/\hat{C}_1$ and $\hat{C}_4/\hat{C}_1$.
We note that the development of a quantitative hydrodynamics framework to describe critical fluctuations is in progress~\cite{Rajagopal:2019xwg,Bluhm:2020mpc,Du:2020bxp} which should eventually be able to provide more robust predictions of the critical point signals in both ordinary and factorial cumulants.


\begin{acknowledgments}

We thank M.I.~Gorenstein, X.~Luo, A.~Rustamov, M.~Stephanov, and N.~Xu for fruitful discussions.
This work received support through the U.S. Department of Energy, 
Office of Science, Office of Nuclear Physics, under contract number 
DE-AC02-05CH11231231, DE-SC001346, and within the framework of the
Beam Energy Scan Theory (BEST) Topical Collaboration. 
V.V. acknowledges the support through the
Feodor Lynen Program of the Alexander von Humboldt
foundation. 
C.S. is also supported in part by the National Science Foundation (NSF) under grant number PHY-2012922.
The computational resources were provided by the LOEWE-CSC.

\end{acknowledgments}

\appendix


\section{Dependence on the switching energy density}
\label{app:esw}

We used a constant value of the switching energy density $\epsilon_{\rm sw} = 0.26$~GeV/fm$^3$ for all collision energies in the main text.
On the other hand, it has been argued that larger values of $\epsilon_{\rm sw}$ may give a better description of bulk observables at higher collision energies~\cite{Oliinychenko:2020znl}.
Thus, here we explore the dependence of the (anti)proton cumulants on the choice of $\epsilon_{\rm sw}$ at $\sNN = 200$~GeV.

\begin{figure}[t]
  \centering
  \includegraphics[width=.49\textwidth]{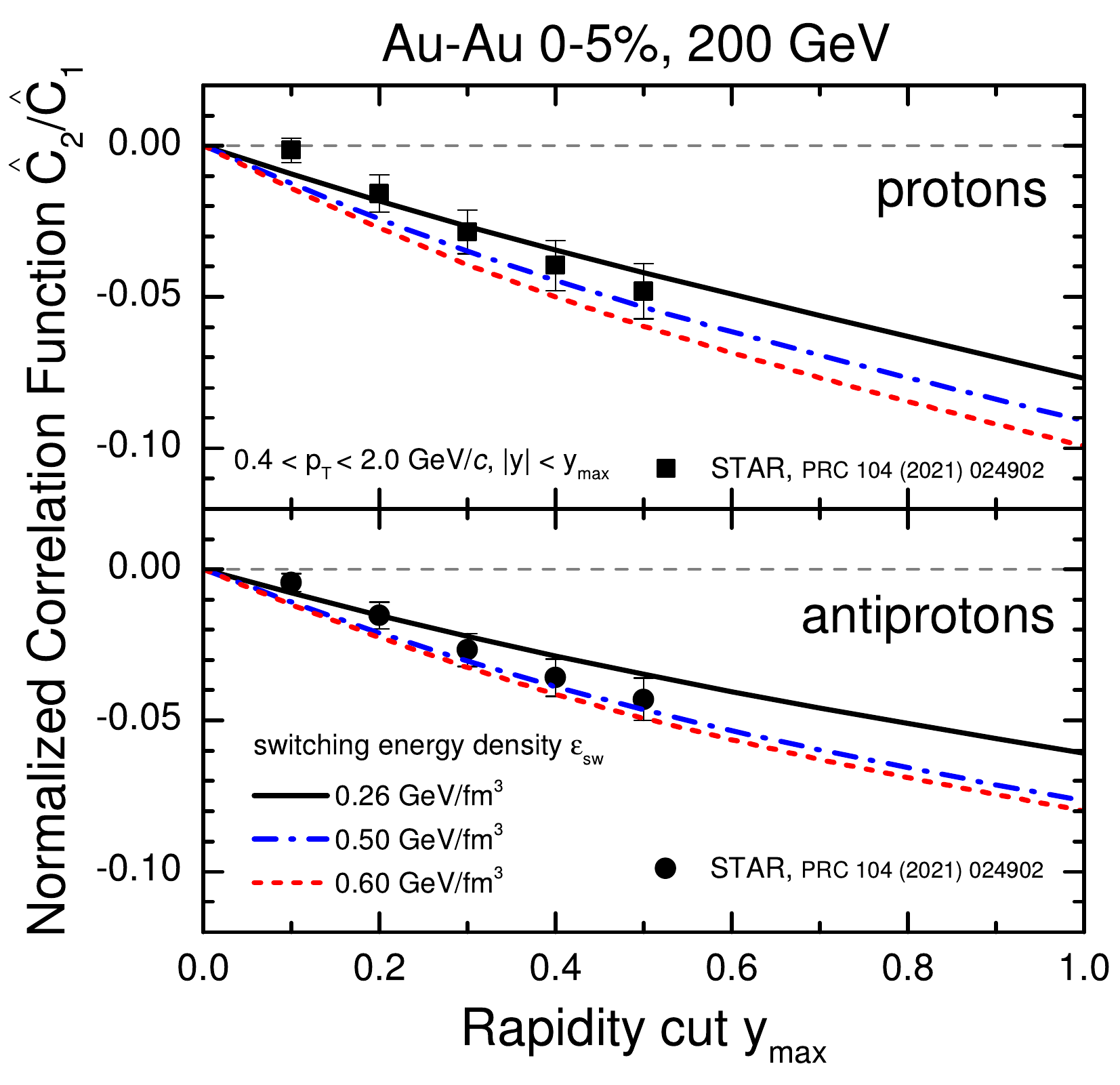}
  \caption{
  Rapidity acceptance dependence of the second normalized factorial cumulant $\hat{C}_2/\hat{C}_1$ of protons~(top) and antiprotons~(bottom) at $\sNN = 200$~GeV calculated from hydrodynamics and compared to the experimental data of the STAR Collaboration~\cite{Abdallah:2021fzj}.
  The calculations incorporate both the excluded volume effect and global baryon conservation and were performed for three different values of the particlization switching energy density $\epsilon_{\rm sw}$: $0.26$~GeV/fm$^3$~(solid black lines), $0.5$~GeV/fm$^3$~(dash-dotted blue lines), 
and $0.6$~GeV/fm$^3$~(dashed red lines).
  }
  \label{fig:esw}
\end{figure}

As the behavior of all the high-order cumulants is primarily driven by the two-particle correlation functions in our model, we focus on the behavior of $\hat{C}_2/\hat{C}_1$ of protons and antiprotons.
Figure~\ref{fig:esw} depicts the rapidity acceptance dependence~($|y| < y_{\rm cut}$) of these quantities for three values of $\epsilon_{\rm sw}$: $0.26$~GeV/fm$^3$~(solid black lines), $0.5$~GeV/fm$^3$~(dash-dotted blue lines), 
and $0.6$~GeV/fm$^3$~(dashed red lines).

The calculations indicate that higher values of $\epsilon_{\rm sw}$ lead to slightly suppressed values of $\hat{C}_2/\hat{C}_1$.
We attribute this effect to a larger role of the excluded volume: particle number densities  increase with  $\epsilon_{\rm sw}$ which leads to stronger effects of repulsion among baryons.
While the resulting sensitivity of $\hat{C}_2/\hat{C}_1$ to the choice of $\epsilon_{\rm sw}$ is not strong, we do observe that $\epsilon_{\rm sw} = 0.5$~GeV/fm$^3$ appears to yield the best agreement with the experimental data of the STAR Collaboration~\cite{Abdallah:2021fzj}.
This observation is in line with Ref.~\cite{Oliinychenko:2020znl}, where the value $\epsilon_{\rm sw} = 0.5$~GeV/fm$^3$ was suggested based on the optimal description of the proton yields at midrapidity.

\section{Validating the analytic results with Monte Carlo}
\label{app:MC}

The cumulants of (net-)(anti-)proton number distribution can be calculated via Monte Carlo sampling of hadrons and resonances at particlization.
These calculations can be used to validate the analytic results obtained in this paper, and to estimate the possible error due to simplifications employed in the analytic procedure, in particular neglecting the difference in the kinematics of nucleons and baryonic resonances, as well neglecting the rescattering in the hadronic phase.

Here we restrict the Monte Carlo calculations to the ideal HRG model.
Monte Carlo sampling of the EV-HRG model is more involved, but can eventually be performed following the subensemble sampler method introduced in Ref.~\cite{Vovchenko:2020kwg}.

The Monte Carlo sampling of hadrons and resonances at Cooper-Frye particlization within the ideal HRG model with exact conservation of baryon number consists of the following steps:
\begin{enumerate}
    \item Mean hadron yields in $4\pi$ are evaluated by summing the grand-canonical ideal HRG model means from each hypersurface element, for each hadron species.
    These means are then scaled such that the total net baryon number is rounded to the nearest integer.
    \item The hadron yields in $4\pi$ for each event are sampled from the Poisson distribution using the pre-computed mean yields.
    \item The event is rejected if the sampled yields violate the exact global baryon number conservation, i.e. if the sampled total net baryon number does not coincide with the expected baryon number computed in the first step.  
    If the sampled yields satisfy the global conservation, we go to the next step.
    \item Momenta and coordinates of each hadron are sampled, independently from all other hadrons.
    To do that first we determine the hypersurface element from which the given hadron is sampled from, this is done via the multinomial distribution where each volume element is weighted by its grand-canonical mean yield for the given hadron species.
    Then the momenta and coordinates of the hadron emitted from the chosen hypersurface element are sampled via the standard procedure.
    \item The chain of all strong, electromagnetic, and weak decays is performed.
\end{enumerate}

\begin{figure}[t]
  \centering
  \includegraphics[width=.49\textwidth]{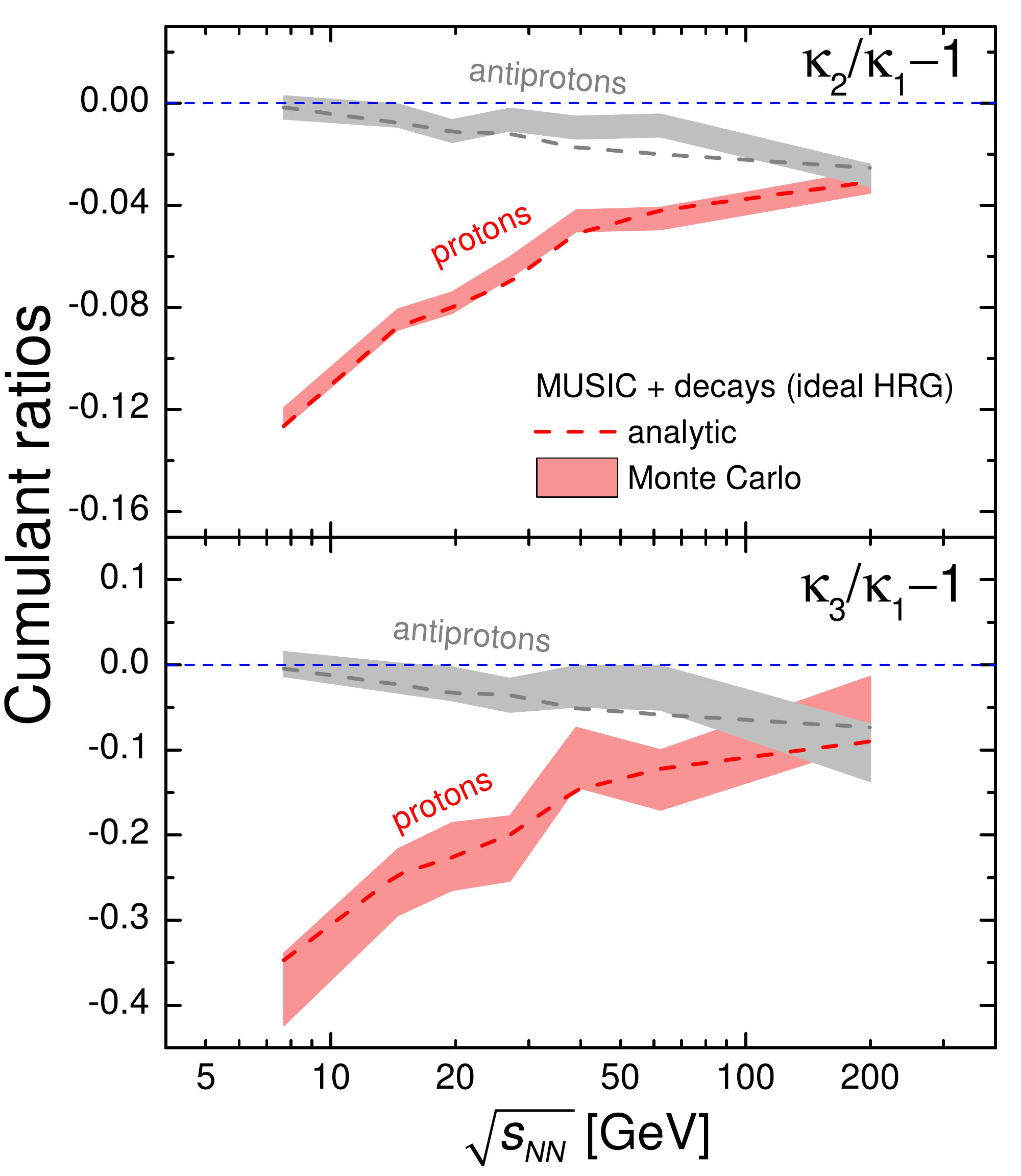}
  \caption{
  Collision energy dependence of scaled (anti)proton cumulants $\kappa_2/\kappa_1 - 1$~(top) and $\kappa_3/\kappa_1 - 1$~(bottom) in 0-5\% Au-Au collisions, evaluated at Cooper-Frye particlization using the ideal HRG model analytically~(dashed lines) and via Monte Carlo sampling~(bands).
  }
  \label{fig:MC}
\end{figure}

We sample $100\,000$ events for each collision energy.
Figure~\ref{fig:MC} depicts the collision energy dependence of the subtracted scaled second and third cumulants $\kappa_2/\kappa_1 - 1$ and $\kappa_3/\kappa_1 - 1$ for protons and antiprotons, evaluated in the STAR momentum acceptance. 
The analytic calculations~(dashed lines) are the same that are shown in Fig.~\ref{fig:KandC} of the main text.
The Monte Carlo results are depicted by the bands. 
The Monte Carlo and analytic results are in good agreement with each other.
This validates the analytic calculations and also indicates that the simplifying assumptions made in the analytic calculation, like neglecting the difference between the masses of nucleons and baryon resonances, have negligible influence on the cumulants.
The Monte Carlo results also validate the accuracy of the generalized subensemble acceptance method of Ref.~\cite{SAM2p0} used to correct the grand-canonical cumulants for baryon number conservation.

Note that although here we explicitly tested only the cumulants up to third order, we expect the analytic calculations to be accurate also for the high-order cumulants. This is due to the fact that both the baryon conservation and excluded volume generate only small multi-particle correlations~(see Fig.~\ref{fig:KandC}), thus the high-order cumulants are mainly determined in this setup by two-particle correlations that we checked to be calculated accurately.
The situation may change if one incorporates effects that generate strong multi-particle correlations like the QCD critical point. In that case the question of accuracy of the analytic calculations for high-order cumulants may have to be revisited.

\section{Exact conservation of electric charge and strangeness}
\label{app:BQS}

In our analysis we have incorporated the effect of exact global baryon conservation but not of other conserved charges like electric charge and strangeness.
While the effect of baryon conservation is expected to be the dominant one, the (anti)proton cumulants are also affected by other conserved charges~\cite{Vovchenko:2020gne,Vovchenko:2020kwg}.
This can be particularly relevant at lower collision energies where protons form a considerable fraction of the total electric charge.
Thus here we evaluate the effect of multiple exactly conserved charges through Monte Carlo sampling within the ideal HRG model.

The sampling algorithm in Appendix~\ref{app:MC} is adjusted in the following way: all events that do not satisfy exact conservation of all three conserved charges are rejected.
The total net strangeness is constrained to be exactly zero in all events while the total electric charge is constrained to reproduce the charge-to-baryon ratio of $Q/B = 0.4$.\footnote{If the total electric charge satisfying the $Q/B = 0.4$ condition is not an integer, it is rounded to the nearest integer.}
To speed up the sampling procedure, we employ the multi-step method of Becattini and Ferroni~\cite{Becattini:2004rq}.
Sampling with multiple conservation laws is more time-consuming than with only baryon number conservation.
We generate about $20\,000$ events for each collision energy and restrict the analysis to the second order moments only.

\begin{figure}[t]
  \centering
  \includegraphics[width=.49\textwidth]{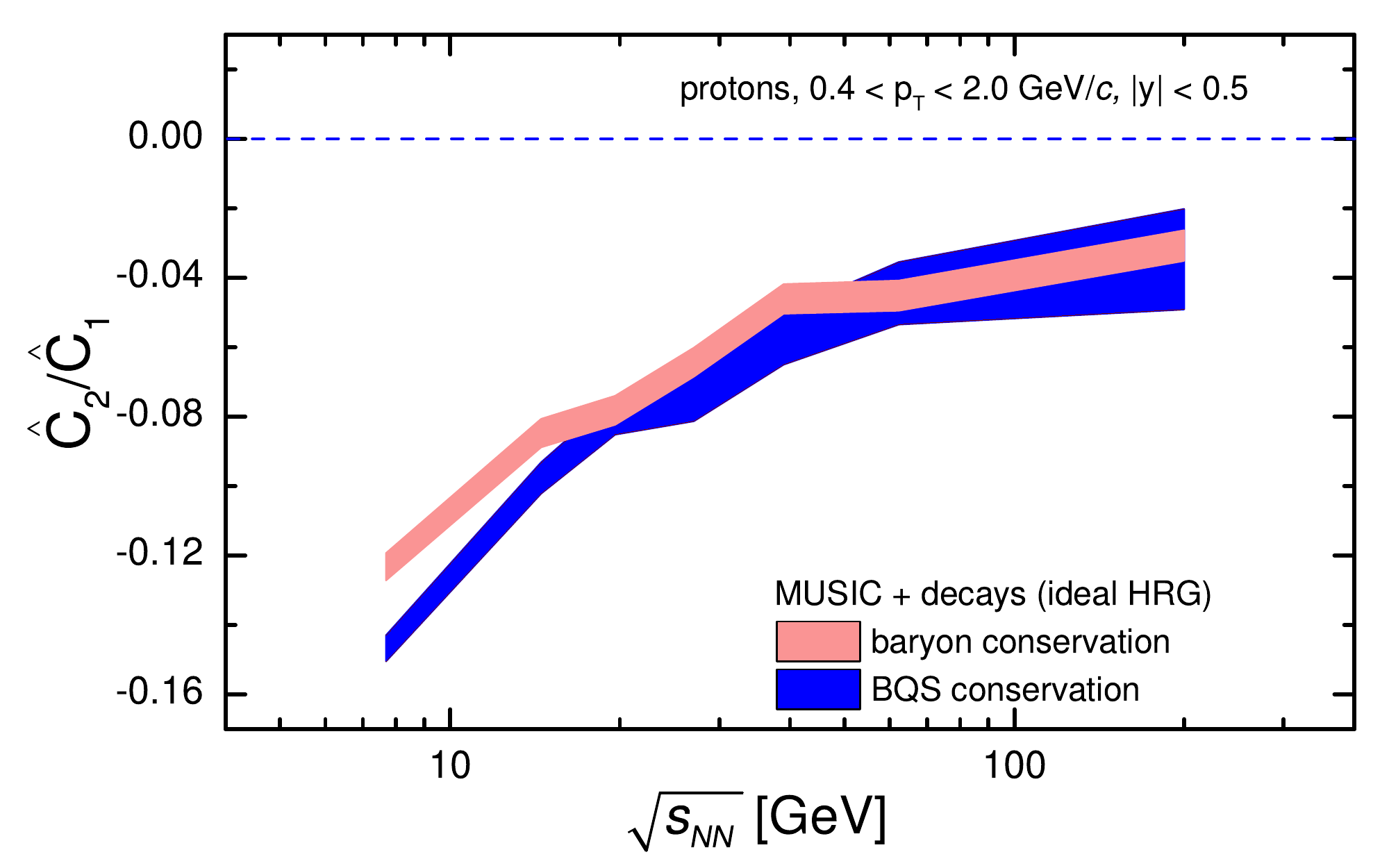}
  \caption{
  Collision energy dependence of the subtracted scaled proton cumulant $\kappa_2/\kappa_1 - 1$ in 0-5\% Au-Au collisions, evaluated at Cooper-Frye particlization using the ideal HRG model via Monte Carlo sampling within the baryon-canonical~(red band) and $BQS$-canonical~(blue band) ensembles.
  }
  \label{fig:BQS}
\end{figure}

The collision energy dependence of the subtracted scaled proton cumulant $\kappa_2/\kappa_1 - 1$ in 0-5\% Au-Au collisions evaluated with the simultaneous conservation of baryon number, electric charge, and strangeness is shown in Figure~\ref{fig:BQS} by the blue band.
The result is compared to the calculations with exact conservation of only baryon number~(red band).
The two calculations agree within statistical errors at $\sNN \gtrsim 20$~GeV, suggesting that the restriction of the global conservation laws to only baryon number might be sufficient in that regime.
At the lower collision energies the conservation of electric charge and strangeness lead to a notable further suppression of the two-particle correlation function of the protons.
Thus, accounting for exact conservation of multiple conserved charges is important for analyses of fluctuations at $\sNN \lesssim 20$~GeV.

\section{Effect of the hadronic phase}
\label{app:UrQMD}

Our analytic calculations neglect rescatterings in the hadronic phase.
The hadronic phase may affect the cumulants in a couple of different ways: (i) it modifies the $p_T$ spectrum of (anti)protons, thus the number of protons in the acceptance may change and (ii) $B\bar{B}$-annihilations may decrease the numbers of protons and antiprotons.
To evaluate the possible effect of the hadronic phase we use the Monte Carlo sampling of the ideal HRG from the previous subsection and, instead of performing the chain of decays, we run the output through the hadronic afterburner UrQMD~\cite{Bass:1998ca,Bleicher:1999xi}.
This is achieved by replacing the step 5 of the Monte Carlo sampling in the previous subsection by the following: all resonances which are not recognized by UrQMD are decayed until only hadrons and resonances recognized by UrQMD are left and then the hadronic phase is simulated by UrQMD via the hadronic afterburner toolkit from~\cite{urqmd-afterburner-toolkit}.

\begin{figure}[t!]
  \centering
  \includegraphics[width=.49\textwidth]{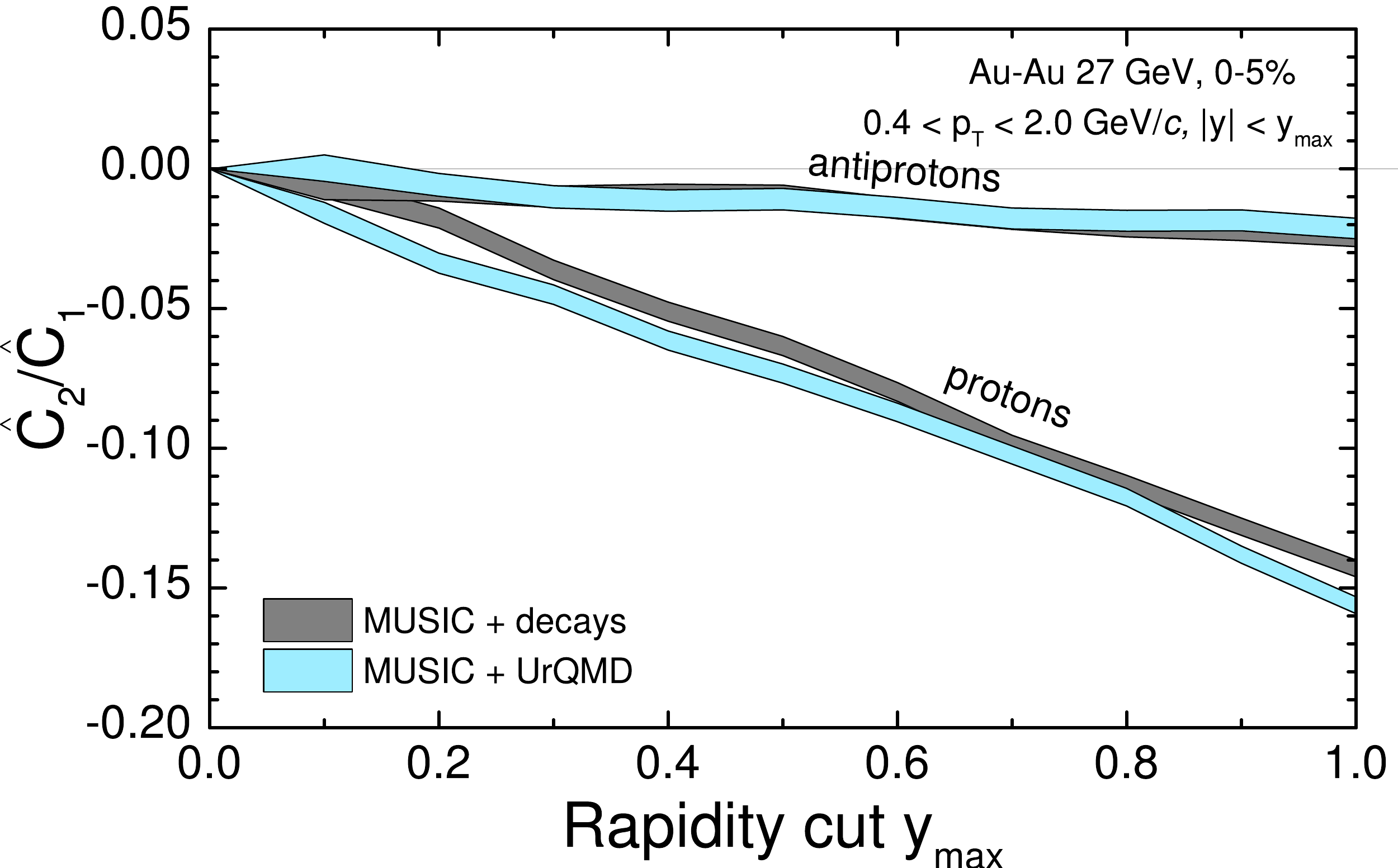}
  \caption{
  Rapidity acceptance dependence of the scaled second factorial cumulant of proton and antiprotons distributions in 0-5\% Au-Au collisions at 
  $\sqrt{s_{NN}} = 27$~GeV calculated in hydro + decays~(gray bands) and hydro + UrQMD~(blue bands) scenarios.
  }
  \label{fig:urqmd}
\end{figure}

We evaluate the effect of hadronic afterburner at $\sNN = 27$~GeV by sampling around $200\,000$ events.
The relevance of the hadronic phase is established by comparing the results with the afterburner~(hydro + UrQMD) to the results without applying the afterburner~(hydro + decays).
Figure~\ref{fig:urqmd} depicts the rapidity acceptance dependence of the second scaled factorial cumulant~$\hat{C}_2/\hat{C}_1$ of protons and antiprotons in the STAR transverse momentum range, $0.4 < p_T < 2.0$~GeV/$c$.
For antiprotons the difference between the two scenarios is within the statistical uncertainty.
For protons the difference is also mild, with indications that~$\hat{C}_2/\hat{C}_1$ is slightly more suppressed when the hadronic phase evolution is included.
This appears to be due to a larger fraction of protons ending up in the STAR acceptance and hence a larger effect of the baryon number conservation.
The hadronic phase leaves the $y_{\rm max}$ dependence of the scaled cumulants essentially unchanged.
Thus, incorporating the time-consuming hadronic afterburner would seem only be necessary for very precise studies of cumulant ratios, at least for $\sNN = 27$~GeV.

\bibliography{BESflucs}


\end{document}